\documentclass[twocolumn]{aastex61}
\pdfoutput=1 %for arXiv submission
\usepackage{amsmath,amstext}
\usepackage[T1]{fontenc}
\usepackage{apjfonts} 
\usepackage[figure,figure*]{hypcap}
\usepackage{graphicx}
\usepackage{subfigure}
\usepackage{soul}
\usepackage{tabu}

\shorttitle{Long Wavelength Dust emission in OMC 2/3}
\shortauthors{Mason et al.}

\begin{document}

\title{Confirmation of Enhanced Long Wavelength Dust Emission in OMC 2/3}
\author[0000-0002-8472-836X]{Brian Mason}
\affiliation{NRAO - National Radio Astronomy Observatory, 520 Edgemont Road, Charlottesville, VA 22903, USA}
\author{Simon Dicker}
\affiliation{Department of Physics and Astronomy, University of Pennsylvania, 209 South 33rd Street, Philadelphia, PA 19104, USA}
\author[0000-0001-7474-6874]{Sarah Sadavoy}
\affiliation{Harvard-Smithsonian Center for Astrophysics (CfA), 60 Garden Street, Cambridge, MA, 02138, USA}
\author{Sara Stanchfield}
\affiliation{Department of Physics and Astronomy, University of Pennsylvania, 209 South 33rd Street, Philadelphia, PA 19104, USA}
\author[0000-0003-3816-5372]{Tony Mroczkowski}
\affiliation{ESO - European Southern Observatory, Karl-Schwarzschild-Str.\ 2, DE-85748 Garching b. M\"unchen, Germany}
\author[0000-0001-5725-0359]{Charles Romero}
\affiliation{Department of Physics and Astronomy, University of Pennsylvania, 209 South 33rd Street, Philadelphia, PA 19104, USA}
\author{Rachel Friesen}
\affiliation{NRAO - National Radio Astronomy Observatory, 520 Edgemont Road, Charlottesville, VA 22903, USA}
\author[0000-0003-0167-0981]{Craig Sarazin}
\affiliation{Department of Astronomy, University of Virginia, 530 McCormick Rd., Charlottesville, VA 22904, USA}
\author[0000-0001-6903-5074]{Jonathan Sievers}
\affiliation{McGill University, 3600 University Street, Montreal, QC H3A 2T8, Canada}
\author[0000-0002-5812-9232]{Thomas Stanke}
\affiliation{ESO - European Southern Observatory, Karl-Schwarzschild-Str.\ 2, DE-85748 Garching b. M\"unchen, Germany}
\author[0000-0002-3169-9761]{Mark Devlin}
\affiliation{Department of Physics and Astronomy, University of Pennsylvania, 209 South 33rd Street, Philadelphia, PA 19104, USA}

\begin{abstract}
Previous continuum observations from the MUSTANG camera on the 
Green Bank Telescope (GBT) of the nearby star-forming filament OMC 2/3 found elevated emission at 
3.3~mm relative to shorter wavelength data.  As a consequence, the inferred dust emissivity
index obtained from modified black body dust spectra was considerably lower than what is typically measured
on $\sim 0.1 \, {\rm pc}$ scales in nearby molecular clouds.   Here we present new observations of OMC 2/3 collected with the MUSTANG-2 camera on the GBT which confirm this elevated emission.
We also present for the first time sensitive 1~cm observations made with the Ka-band receiver on the GBT which 
also show higher than expected emission.  
We use these observations--- 
along with {\it Herschel}, JCMT, Mambo, and GISMO data--- 
to assemble spectral energy distributions (SEDs) of a variety of structures in OMC 2/3
spanning the range  $160 \, {\rm \mu m}$ to $1 \, {\rm cm}$.  
The data at 2~mm and shorter are generally consistent with a 
modified black body spectrum and a single value of $\beta \sim 1.6$.
The 3~mm and 1~cm data, however, lie well above such an SED.  The spectrum
of the long wavelength excess is inconsistent with both free-free emission
and standard ``Spinning Dust'' models for Anomalous Microwave Emission (AME).
The 3~mm and 1~cm data could be explained by a flatter dust emissivity 
at wavelengths shorter than 2~mm, potentially in concert with AME in some regions.

\end{abstract}


\section{Introduction}

OMC 2/3 is the richest known star-forming filament within 500 pc and
has been studied extensively at millimeter, sub-millimeter, and
infrared wavelengths \citep[e.g.][]{johnstone1999,peterson2005,nutter2007,davis2009,sadavoy2010,herschel2015,megeath2016,salji2015}.  These studies generally aim to map the
distribution of star formation within OMC 2/3 to determine how the dense core or young star populations relate to the dynamics of the filament, or to determine the characteristics
of the filament itself.  Thermal
dust emission is particularly useful as a tracer of mass in
filaments and star-forming regions more generally. 
Dust is typically assumed to emit  thermal radiation with a modified
black body spectrum:
\begin{equation}
I_{\nu,dust}    \propto \frac{\nu^{(3 +\beta)}}{exp(h\nu/k T_d)-1}.
\end{equation}
Here $T_d$ is the dust grain temperature and the dust emissivity index $\beta$ is determined by physical
properties of the dust such as composition and grain size distribution. This model is generally seen
to be an excellent description of dust emission at millimeter and sub-millimeter
wavelengths, with typical values $1.5 < \beta < 2.5$ on filament to molecular cloud
scales \citep[e.g.][]{goldsmith1997,sadavoy2013}.
Observations 
of diffuse thermal dust emission at long millimeter wavelengths 
can be  challenging due to the relative faintness of the emission, but 
are appealing because mass determinations from them are less affected by optical
depth considerations than those obtained from shorter wavelength data.  In addition,
their greater spectral leverage gives greater sensitivity to the physical 
properties of dust grains, potentially revealing new physical information. 
These long wavelength observations have become readily feasible by advances
in instrumentation, such as ALMA, and focal plane arrays on large single dish telescopes.

With this in mind \citet{schnee2014} combined observations of
OMC 2/3 at $\lambda=3.3~{\rm mm}$ with MUSTANG on the Robert C. Byrd Green Bank Telescope (GBT) with observations at $\lambda = 1.2~{\rm  mm}$ from MAMBO at the IRAM 30m telescope to study the dust emissivity index in the filament on $\lesssim 0.1 \, {\rm pc}$ scales.  \citet[][hereafter
  S14]{schnee2014} found 
surprisingly high 3~mm emission toward regions dominated by thermal
dust emission at shorter wavelength. S14 tentatively attributed this
to grain growth in the filaments resulting in a lower value of $\beta
\approx 0.9$.  
Subsequent
analysis \citep{sadavoy2016} used {\it Herschel} sub-mm telescope data at $160$ - $500 \, {\rm \mu m}$, as well as 2~mm data from
GISMO on the IRAM 30m.  
 \citet[][hereafter S16]{sadavoy2016} measured $\beta$ values of $\sim 1.7$ on $0.1 \, {\rm pc}$ scales. These values are more consistent with the typical indices found with {\it Planck} \citep{planckDust2015} for molecular clouds and suggests that the dust grains on OMC 2/3 are not unusually large.  S16 showed that the $\beta$ values in S14 may have been low due to a break in the dust SED at $\lambda > {\rm 2 mm}$, such that the 3mm emission appeared to be elevated.

In the course of commissioning the new MUSTANG-2 camera
on the GBT we obtained a 60 minute
observation of OMC 2/3.  The results of these observations, as well as new
31 GHz continuum data from the GBT, are presented here. MUSTANG-2 is more sensitive than
the original MUSTANG camera. Of particular significance to the
interpretation of these data is the fact that it has a much larger
Field of View (FOV; $4\arcmin\!.25$ {\it vs.} $42\arcsec$). The larger FOV enables
much more straightforward reconstruction of extended, diffuse signals
and thus provides a valuable cross-check on the results of S14, the
achievement of which required a sophisticated, iterative
reconstruction algorithm.  As a point of comparison MUSTANG-2 readily
measures spatial scales 6 times larger than can currently be measured by ALMA
in the continuum at these wavelengths.

The structure of this paper is as follows. Section~\ref{sec:observations} presents our new $3.3 \, {\rm mm}$ and $1 \, {\rm cm}$ observations of OMC 2/3. Section~\ref{sec:analysis} describes the analysis leading to multi-wavelength SEDs for a variety of structures in the region, and Section~\ref{sec:interpretation} considers several possible physical interpretations of them. Finally, Section~\ref{sec:conclusions} reviews and presents our conclusions. Except as stated all error bars indicate $1\sigma$ ($68\%$ confidence) uncertainties on the quantity of interest.

\section{Observations}
\label{sec:observations}

\subsection{MUSTANG-2}

MUSTANG-2 is a 215 pixel feedhorn coupled Transition Edge Sensor (TES)
bolometer array with a bandpass of $75-105 \, {\rm GHz}$. The receiver
is cooled with a pulse tube and closed cycle Helium 4/Helium 3
refrigerator, which cools the array to $300 \, {\rm mK}$.  More  information on the MUSTANG-2 receiver can be found in \citet{dicker2014} and \citet{stanchfield2016}.

Observations of OMC 2/3 were
acquired in two observing sessions in 2016 December. The data comprise eight individual scans, each approximately 8 minutes in duration, centered on five pointing centers chosen to cover the ``Integral Shaped Filament'' in OMC 2/3.  Each scan covers a circular area approximately $6^\prime\!.5$ in diameter.
The data were calibrated and imaged by a suite of software tools  developed by the instrument team in IDL.  The time-ordered data were visually inspected for data quality and a small fraction of detectors manually flagged, supplementing automatic detector flags inferred and applied by the calibration software.   Local pointing corrections were derived and applied at the map-making stage using bracketing observations of the nearby pointing source J0530+135. Flux density calibration was performed using an observation of 3C84, one of the sources that ALMA regularly monitors \citep{Fomalont2014,vanKempen2014}.
While 3C84 shows significant long-term time variability, the ALMA 3~mm data show it is stable to within $\sim 5\%$ (peak to peak) in the 2 months leading up to and encompassing our MUSTANG-2 observations.  We assume an interpolated flux density of $23.8 \, {\rm Jy}$  at $3.3 \, {\rm mm}$ based on the ALMA $91.5 \, {\rm GHz}$ measurements.

The telescope beam was mapped using both 3C84 and the secondary calibration source, J0530+135.  The main beam has a full width at half-maximum (FWHM) between $9\arcsec\!\!.1$ and $9\arcsec\!\!.6$, varying
with time depending on the thermal state of the antenna. Beam size variations larger than this are removed by periodic ``Out of Focus Holography'' measurements \citep{bojanoof}.
An error beam is evident with a peak normalization
$\sim 4\%$ of the main beam and a FWHM $\sim 30''$. The ratio of main
beam area to error beam area is seen to be $2.42$ in the stacked beam
maps from 0530+135, and $2.35$ in the beam maps using 3C84. The
characteristics of the beam determined from individual, quick maps of
these sources are consistent with the beam characteristics determined
from stacked measurements, indicating that telescope pointing has been
well corrected in the maps.

\begin{figure}
\centering
\includegraphics[width=\columnwidth]{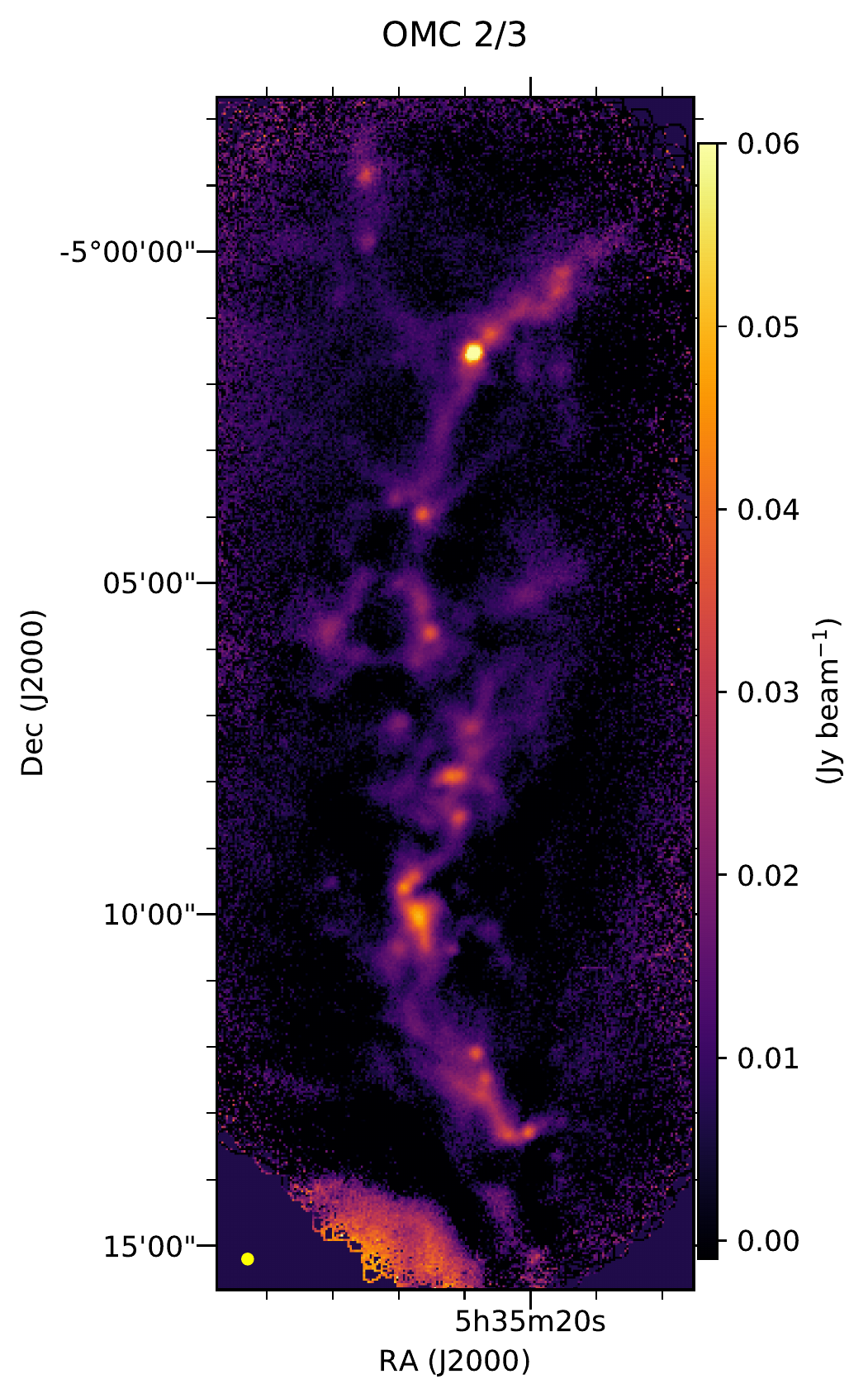}
\caption{The $3.3$~mm MUSTANG-2 map at full angular resolution. The HII region M43 can be seen toward the lower edge of the map. The 9\arcsec\ MUSTANG-2 beam is depicted in yellow in the lower left corner.}
\label{fig:newOldPix}
\end{figure}

Figure~\ref{fig:newOldPix} shows the MUSTANG-2 map of OMC 2/3.  
The shape and intensity of the features seen in this map correspond very well with those in the MUSTANG-1 map presented in S14.  
Despite being made in a much shorter time (40 minutes on source {\it vs.} 14 hrs on source) the new map has slightly lower noise than the old map ($0.8 \, {\rm mJy \, bm^{-1}}$ {\it vs.} $1.0 \, {\rm mJy \, bm^{-1}}$). Due to the larger field of view ($4'.25$ {\it vs.} $42''$),  MUSTANG-2 data  allow the large-scale features to be reconstructed more easily, whereas S14 needed to use a more complex, iterative imaging algorithm to do so.  This comes
about because an important step in imaging ground-based, millimeter continuum cameras is removing atmospheric emission. Generally this has the effect of filtering out spatial scales larger than the camera's instantaneous field of view. S14 present details on the approach used to reconstruct larger scales from the original MUSTANG observation, and \citet{romero2019}  present a quantitative evaluation of the transfer function of the IDL analysis pipeline used in this work. 

We performed a quantitative comparison between the maps by gridding them on the same set of pixels, and place them on a common surface brightness scale by correcting for the modest difference in beam size between the images which arises from the differing aperture illumination patterns of MUSTANG and MUSTANG-2. Selecting all ($3,557$) pixels that are above $3\sigma$ in both maps, we find a median ratio of old to new pixel values of $0.95$ and a mean ratio of $1.0$.  A least squares fit for the slope old pixel values as a function of new pixel values yields a slope of $1.03$.  The 3~mm maps of OMC 2/3 are thus robustly in agreement 
in spite of having been made with a different instrument, as well as different calibrations  and image reconstruction techniques.

An HII region is visible at the southern edge of the MUSTANG-2 map.
This HII region, M43, is generated by the early
B-star HD 37061 / Nu Ori \citep{Simon-Diaz2011}. It is
also evident in the GISMO ($2.1 \, {\rm mm}$) map (S16) and as a cavity at $850 \, {\rm \mu m}$ from
SCUBA-2 \citep{salji2015}.  Inspection of these
maps along with the 21~cm continuum image \citep[NVSS,][]{condon1998}
reveals that the 3~mm and 2~mm maps, while probably dominated by
Free-Free emission toward the HII region, have features that are not seen at 21~cm. This
suggests some amount of dust contamination, a conclusion supported by
measurements in the far-IR \citep{smith1987}. These features are particularly noticable
at the outer edge of M43.

%make this analysis challenging at best. First, the 21~cm data may not
%be at a high enough frequency that the spectrum is optically then
%\citep{thum1978}, resulting in an uncertain radio spectrum.  Second,
%the three maps resolutions are substantially different: $45''$ (FWHM)
%for NVSS; $9''$ for MUSTANG-2; and $21''$ for GISMO. Combined with the
%fact that M43 is truncated by the map edges, this makes a comparison
%difficult. Finally, detailed 

%\begin{figure*}
%\begin{center}
%\includegraphics[width=5in]{nvss_m2smo_gismo_m43zoom_newAnnot}
%\end{center}
%\caption{Detail showing the M43 HII region at $1.4 \, {\rm GHz}$ (NVSS
%  - left); 3~mm (MUSTANG - center); and 2mm (GISMO - right).  Markers
%  and slices are at the same astronomical locations in each image.}
%\label{fig:m43}
%\end{figure*}
%\clearpage

The MUSTANG-2 map of OMC-2/3 is publicly available  
on the Harvard Dataverse ({\tt doi:10.7910/DVN/EWN4RK}).

\subsection{GBT 31 GHz Observations}
\label{sec:1cm}

After confirming the enhanced 3~mm emission originally seen with MUSTANG, the MUSTANG-2 team obtained Director's Discretionary Time on the GBT at Ka Band (program code AGBT18A-446).  The observations used the Caltech Continuum Backend to perform sensitive, wide-band continuum beam switching, and the data were
calibrated and reduced using standard procedures described in \citet{mason2009}. These observations, unlike the others reported here,
consist of single-pointing photometric measurements targeting selected regions. Each measurement also has two ``off'' or reference positions offset in azimuth which are generated by the combination of electronic beam switching and telescope nodding, as described in \citet{mason2009}.  Photometric data were collected targeting the location with the highest 3~mm brightness in each of the 24 regions or ``slices'' studied in detail (see \S~\ref{subsec:slices}) on 2018 March 5. The data were calibrated with respect to 3C138, using VLA Ka Band data obtained on 2018 February 26 as part of VLA polarization calibration (service mode) observations.  The VLA observations of 3C138, calibrated relative to 3C286 using the standard VLA calibration scale, indicate a flux density of $0.86 \, {\rm Jy}$ for 3C138.  While the GBT Ka-band receiver provides four separate, $3.5 \, {\rm GHz}$ channels covering the $14 \, {\rm GHz}$ receiver band, the four channels have been averaged for purposes of this analysis in order to avoid over-weighting the 1~cm spectral information

Astronomical signal in the reference positions will generate negative offset in the photometric measurement. Only one measurement--- that of Slice 2--- was clearly affected by contamination in the reference position. While we cannot rule out low level contamination in other positions, the reference positions were separated from the on-source positions primarily in Right Ascension, {\it i.e.} off the filament. Furthermore the $78''$ separation between the main position and each of the reference positions is comparable to the typical $1' - 1'.5$ half-lengths of the slices used to extract photometry at shorter wavelengths (\S~\ref{subsec:slices}), such that all the points in a given SED should have comparable reference levels.  Significant emission was clearly detected in all but one of the 24 pointing positions. The exception is the previously mentioned Slice 2, which has been excluded from all further analysis.

%Note that any residual contamination would spuriously suppress, rather than elevate, the measured 30 GHz photometry.

There are a number of very young stellar systems in OMC 2/3 which can give rise to thermal bremmstrahlung emission.   \citet[][hereafter R99]{Reipurth1999} mapped this region at $8''$ resolution using the VLA at $8.3 \, {\rm GHz}$ and detected 14 sources.  We have used this source list, converted into a map, assuming an optically thin free-free spectrum ($\propto \nu^{-0.1}$), and smoothing  to the $24''$ GBT 31 GHz resolution, to correct our GBT 1cm data for free-free emission.   Eight regions  had non-zero corrections. Of these, the corrections to the measured flux densities of four regions were $> 15\%$: slice 10 ($25\%$), slice 12 ($48\%$), slice 14 ($52\%$), and slice 18 ($34\%$). The Ka band measurements and their free-free corrections are summarized in Table~\ref{tbl:1cmdata}.

% going by rule #8 of https://journals.aas.org/manuscript-preparation/#style:
% Right ascension and declination in text and equations are given in the form: 3h25m8s.15, 90°26’14 5″.

\begin{table}[]
    \centering
    \begin{tabular}{c|c|c|c|c}
Slice & R.A. & Dec. & $S_{31}$ & $f\!f_{corr}$ \\ 
     & (J2000) & (J2000) & (mJy bm$^{-1}$) & (mJy bm$^{-1}$) \\ \hline
1 & $05^\mathrm{h}35^\mathrm{m}30^\mathrm{s}$ &  $-04^\circ58^{\prime}49\arcsec$      &  $ 1.32 \pm     0.12 $  &   -    \\
2 & $05^\mathrm{h}35^\mathrm{m}28^\mathrm{s}$ &  $-04^\circ59^{\prime}44\arcsec$     &  $-0.64 \pm     0.11 $  &    -    \\
3 & $05^\mathrm{h}35^\mathrm{m}14^\mathrm{s}$ &  $-04^\circ59^{\prime}32\arcsec$     &  $ 1.31 \pm     0.12 $  &  -   \\  % typo: 3->4 ffcorr
4  & $05^\mathrm{h}35^\mathrm{m}18^\mathrm{s}$ &  $-05^\circ00^{\prime}22\arcsec$     &  $ 3.01 \pm     0.12 $  &    $0.10$    \\ % fix and rerun using tweaked sliceNod. ka sed file. just like region 7.
5 & $05^\mathrm{h}35^\mathrm{m}19^\mathrm{s}$ &  $-05^\circ00^{\prime}43\arcsec$     &  $ 2.56 \pm     0.12 $  &   -    \\
6/7 & $05^\mathrm{h}35^\mathrm{m}23^\mathrm{s}$ &  $-05^\circ01^{\prime}33\arcsec$     &  $ 6.11 \pm     0.12 $  &   $0.12$   \\
8 & $05^\mathrm{h}35^\mathrm{m}25^\mathrm{s}$ &  $-05^\circ02^{\prime}31\arcsec$     &  $ 1.25 \pm     0.12 $  &   -    \\
9 & $05^\mathrm{h}35^\mathrm{m}26^\mathrm{s}$ &  $-05^\circ03^{\prime}07\arcsec$     &  $ 0.38 \pm     0.12 $  &    -    \\
10 & $05^\mathrm{h}35^\mathrm{m}26^\mathrm{s}$ &  $-05^\circ03^{\prime}55\arcsec$       &  $ 2.02 \pm     0.12$  &   $0.5$    \\
11 & $05^\mathrm{h}35^\mathrm{m}19^\mathrm{s}$ &  $-05^\circ05^{\prime}11\arcsec$     &  $ 2.11 \pm     0.12$  &   -    \\
12 & $05^\mathrm{h}35^\mathrm{m}26^\mathrm{s}$ &  $-05^\circ05^{\prime}43\arcsec$      &  $ 1.45 \pm     0.12$  &   $0.7$    \\
13 & $05^\mathrm{h}35^\mathrm{m}17^\mathrm{s}$ &  $-05^\circ06^{\prime}03\arcsec$      &  $ 1.21 \pm     0.12$  &   -    \\
14 & $05^\mathrm{h}35^\mathrm{m}24^\mathrm{s}$ &  $-05^\circ07^{\prime}01\arcsec$      &  $ 1.91 \pm     0.12$  &   $1.0$    \\
15 & $05^\mathrm{h}35^\mathrm{m}25^\mathrm{s}$ &  $-05^\circ07^{\prime}56\arcsec$      &  $ 2.70 \pm     0.12$  &   $0.24$   \\
16 & $05^\mathrm{h}35^\mathrm{m}35^\mathrm{s}$ &  $-05^\circ08^{\prime}21\arcsec$     &  $ 0.52 \pm     0.12$  &    -    \\
17 & $05^\mathrm{h}35^\mathrm{m}26^\mathrm{s}$ &  $-05^\circ09^{\prime}02\arcsec$      &  $ 1.27 \pm     0.16$  &   -    \\
18 & $05^\mathrm{h}35^\mathrm{m}27^\mathrm{s}$ &  $-05^\circ09^{\prime}33\arcsec$      &  $ 5.00 \pm     0.16$  &   $1.7$    \\
19 & $05^\mathrm{h}35^\mathrm{m}27^\mathrm{s}$ &  $-05^\circ09^{\prime}56\arcsec$      &  $ 5.34 \pm     0.16$  &   $0.7$    \\
20 & $05^\mathrm{h}35^\mathrm{m}26^\mathrm{s}$ &  $-05^\circ10^{\prime}57\arcsec$     &  $ 0.66 \pm     0.16$  &    -    \\
21 & $05^\mathrm{h}35^\mathrm{m}26^\mathrm{s}$ &  $-05^\circ11^{\prime}32\arcsec$      &  $ 1.34 \pm     0.16$  &   -    \\
22 & $05^\mathrm{h}35^\mathrm{m}23^\mathrm{s}$ &  $-05^\circ12^{\prime}07\arcsec$      &  $ 3.28 \pm     0.16$  &   -    \\
23 & $05^\mathrm{h}35^\mathrm{m}23^\mathrm{s}$ &  $-05^\circ12^{\prime}38\arcsec$      &  $ 1.46 \pm     0.16$  &   -    \\
24 & $05^\mathrm{h}35^\mathrm{m}20^\mathrm{s}$ &  $-05^\circ13^{\prime}19\arcsec$      &  $ 1.70 \pm     0.16$  &   -    \\
    \end{tabular}
    \caption{Positions, flux densities, and free-free corrections ($f\!f_{corr}$), if applicable, for the  GBT 31 GHz measurements of 24 regions in OMC 2/3. Note that Slice 2 was not used in the analysis and interpretation due to contamination in one of the beamswitch reference positions.}
    \label{tbl:1cmdata}
\end{table}

\subsection{Archival Data}
\label{subsec:otherdata}

\begin{figure*}
\centering

\gridline{\includegraphics[scale=0.5]{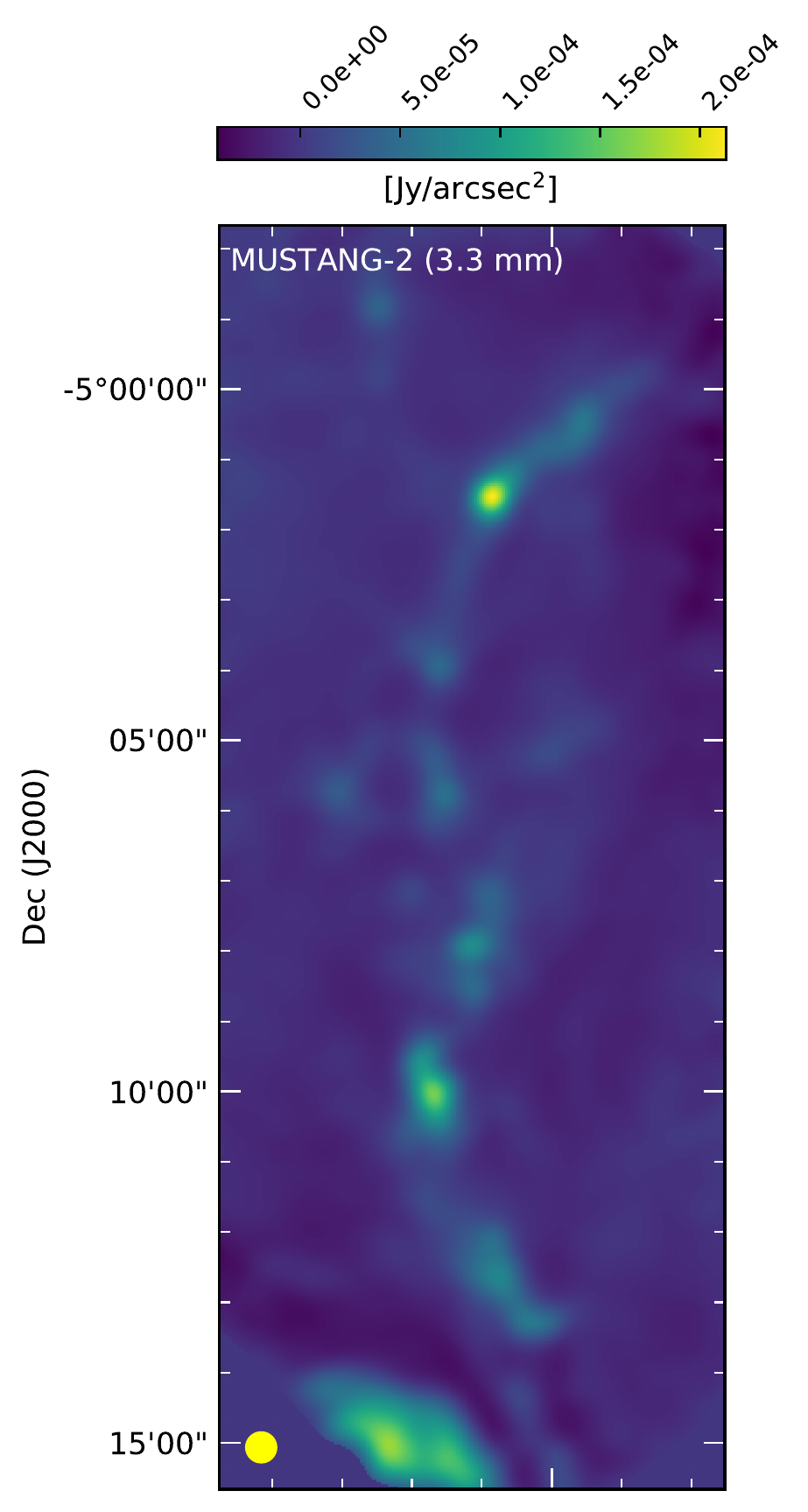}
\includegraphics[scale=0.5]{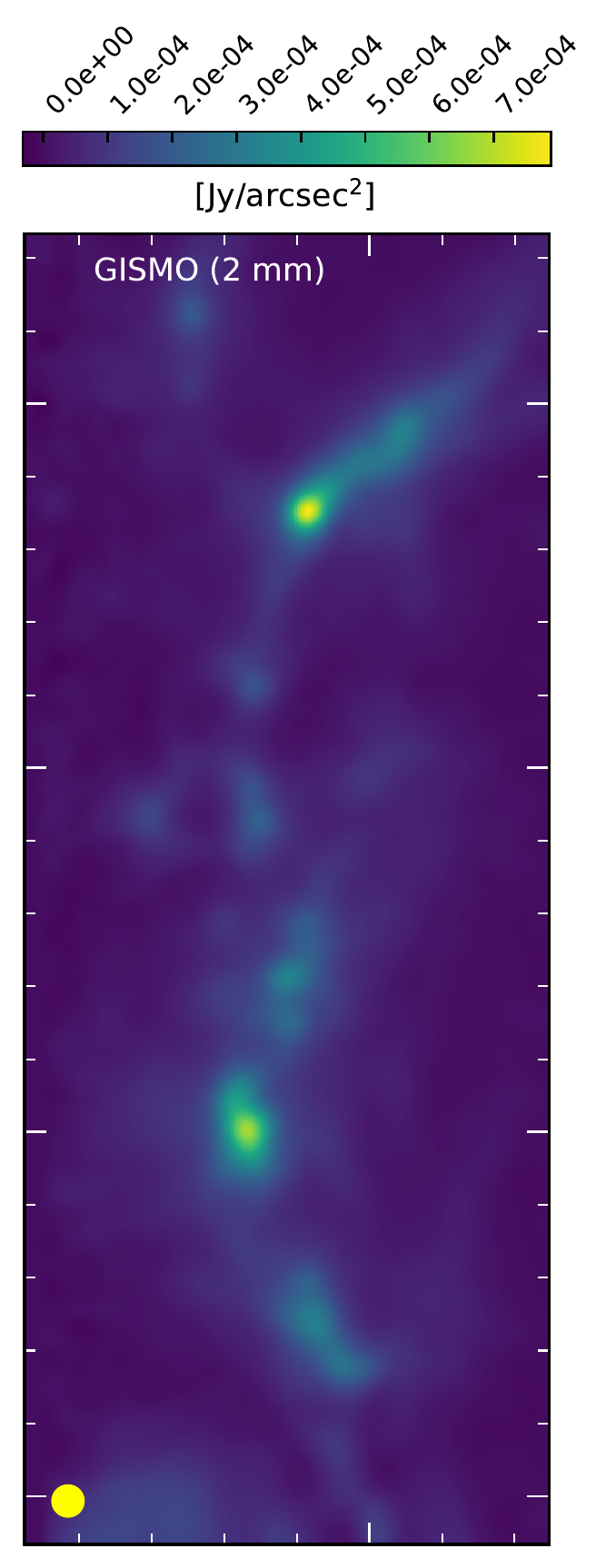}
\includegraphics[scale=0.5]{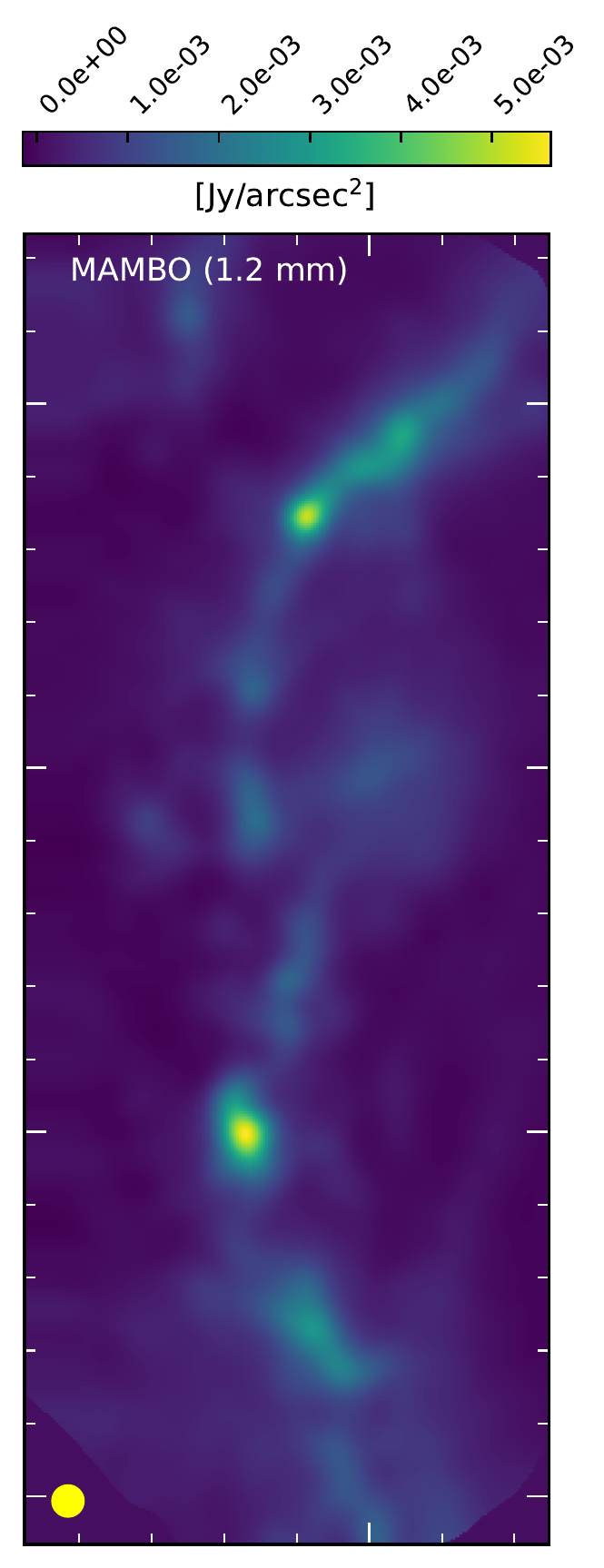}
\includegraphics[scale=0.5]{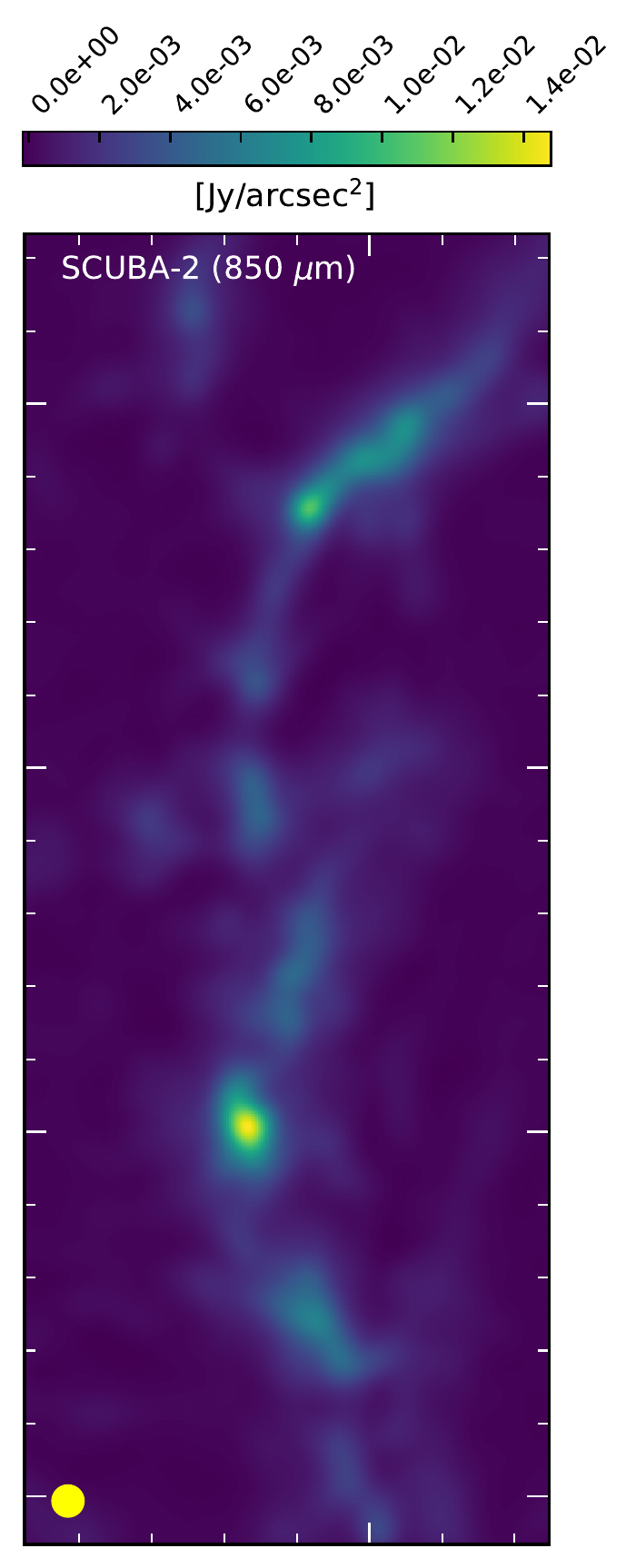}
}
\gridline{\includegraphics[scale=0.5]{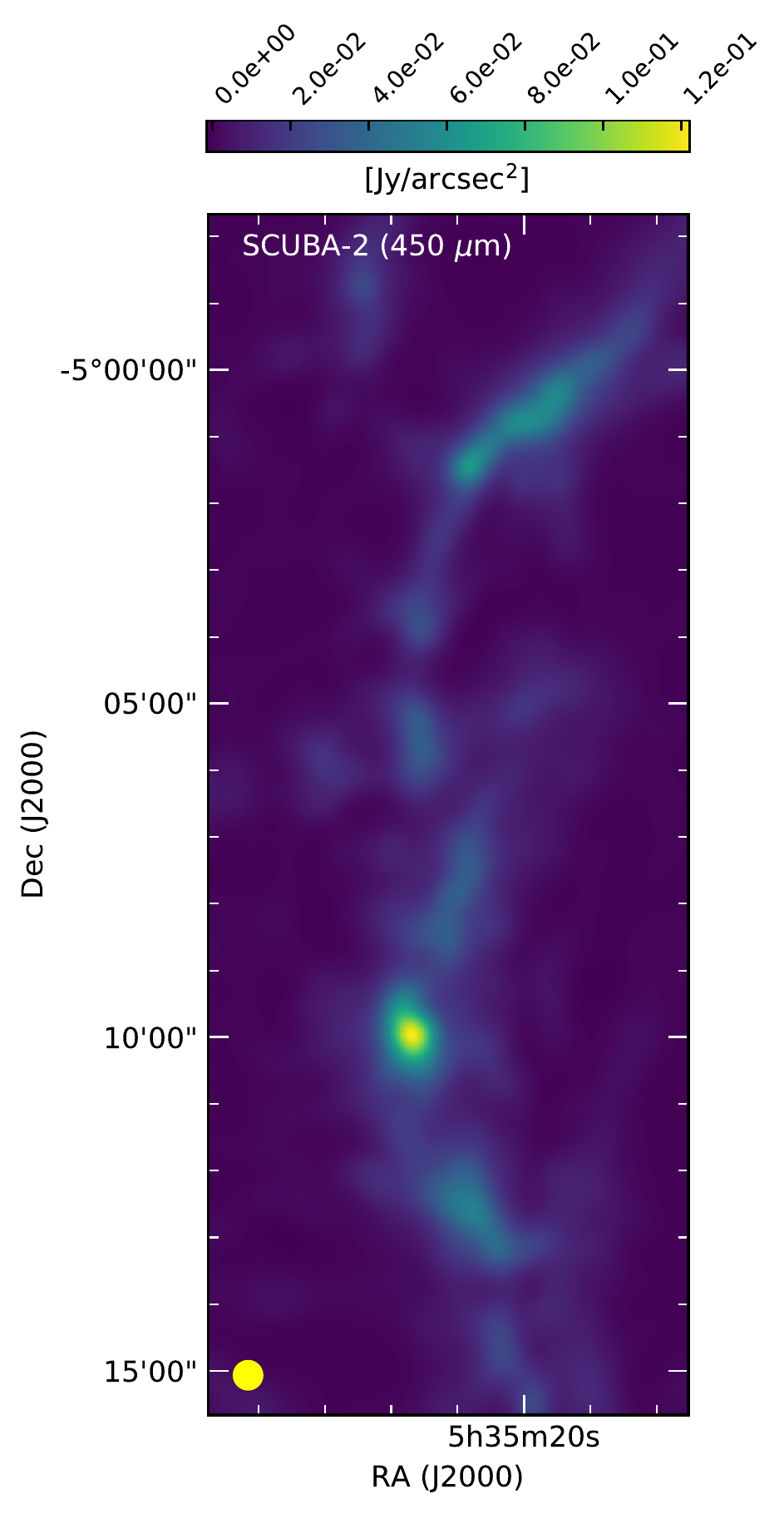}
\includegraphics[scale=0.5]{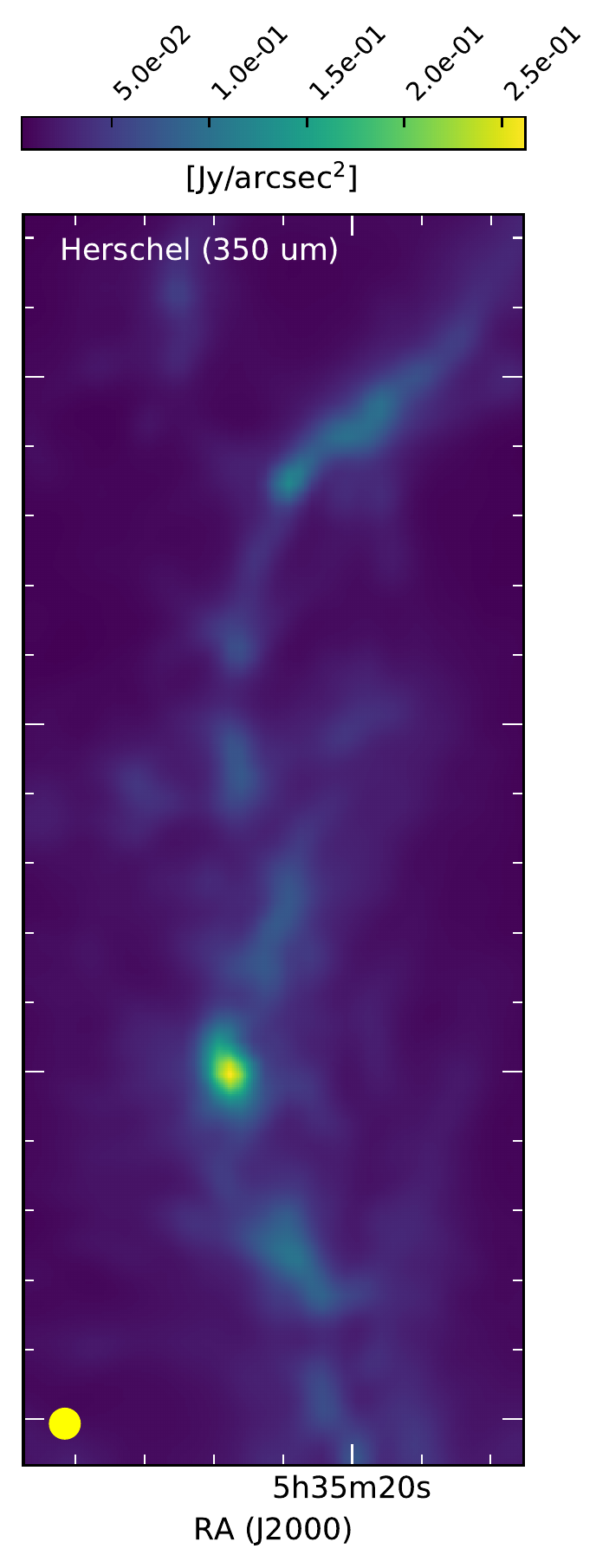}
\includegraphics[scale=0.5]{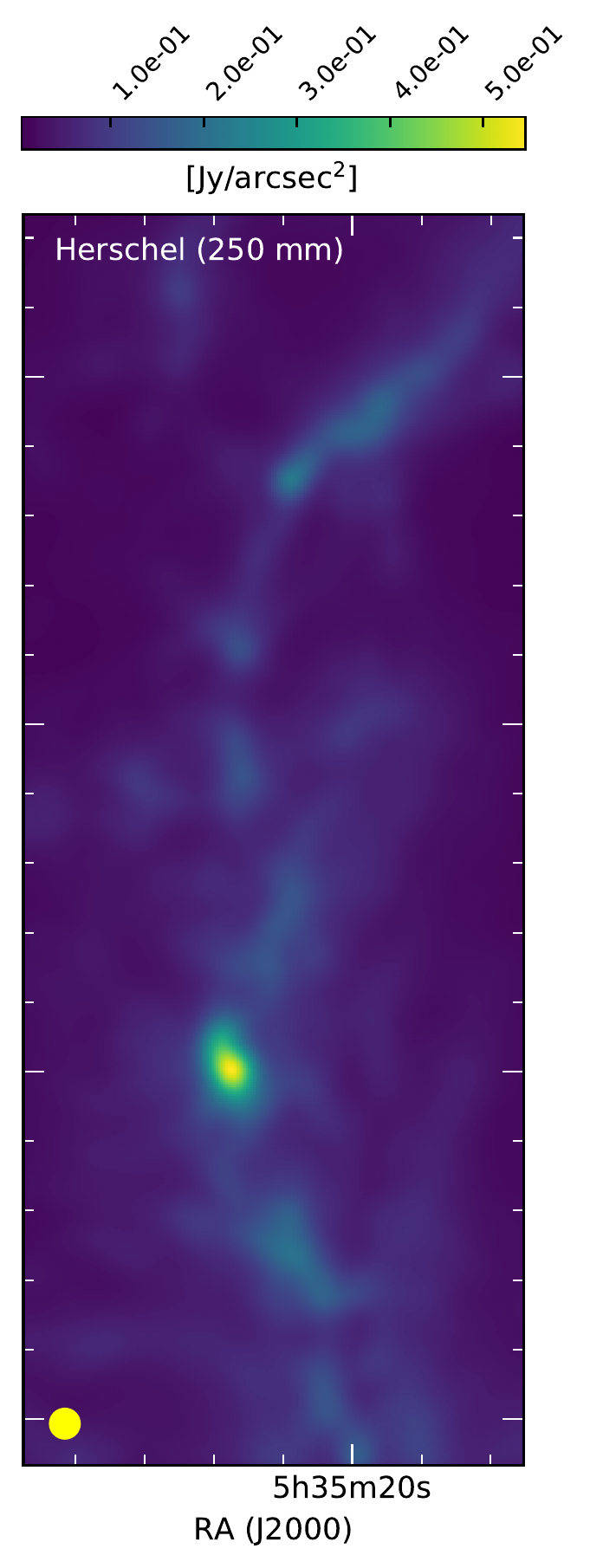}
\includegraphics[scale=0.5]{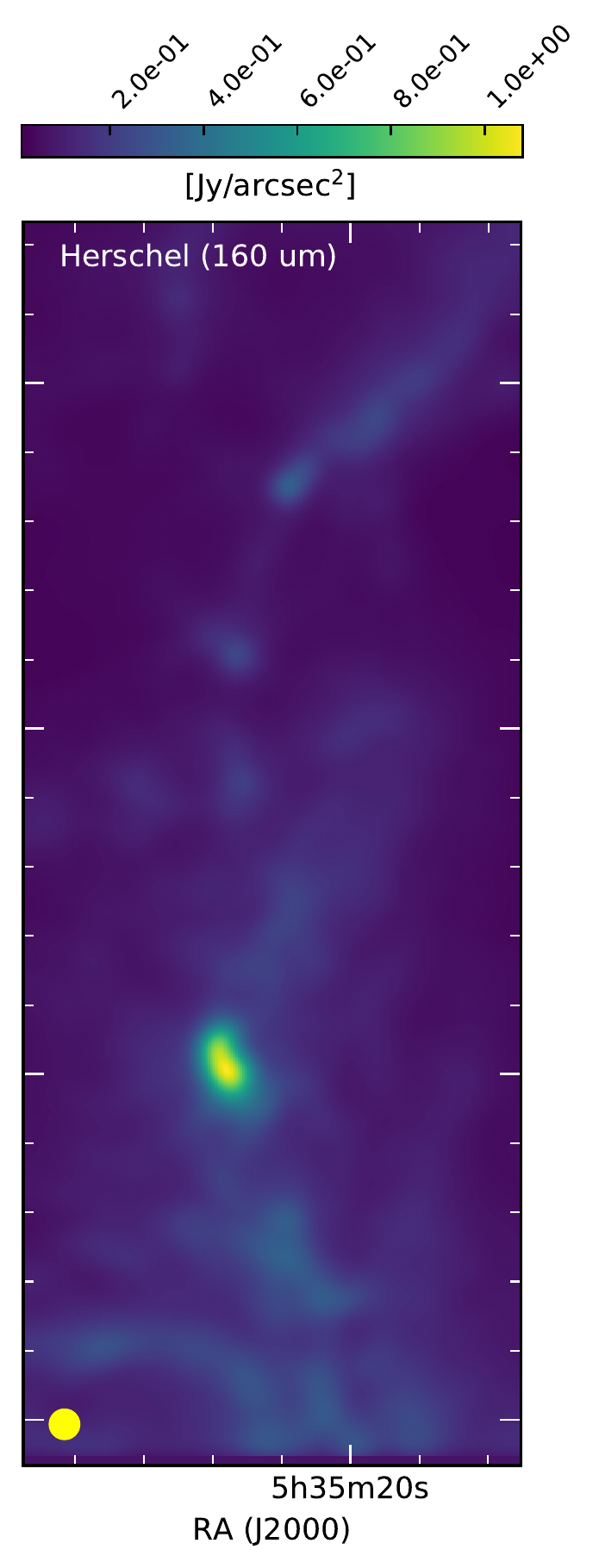}
}
\caption{MUSTANG-2, GISMO, MAMBO, SCUBA-2, and Herschel maps used in this analysis. All maps have been smoothed to $25''$, which is the native resolution of the Herschel $350 \, {\rm \mu m}$ map.
}
\label{fig:allDataWslices}
\end{figure*}

In order to interpret these data we bring to bear the $1.2 \, {\rm mm}$ MAMBO
map presented in S14 along with the multi-wavelength dataset presented
by S16: 2mm data from GISMO in the IRAM 30m; $450 \, {\rm \mu m}$ and $850 \, {\rm \mu m}$ SCUBA-2 maps from the JCMT Gould Belt Survey \citep{mairs2016};
and Herschel $160 \, {\rm \mu m} - 350 \, {\rm \mu m}$ maps \citep{herschel2015}.   These datasets, convolved to a $25'' (FWHM)$ common resolution, are shown in Fig.~\ref{fig:allDataWslices}.  Herschel $500 \, {\rm \mu m}$ data were not used because they were very close in wavelength to the SCUBA-2 $450 \, {\rm \mu m}$ data but had worse angular resolution.

% actually the project compared to was
% 2013.1.00662.S PI = Diego Mardones
OMC 2/3 has also been observed by ALMA at a similar frequency range to the observations we report here under the auspices of project {\tt 2013.1.00662.S} (PI: D. Mardones). These data covered OMC-2 and OMC-3 over a narrow ($91.2$ - $91.7$ GHz) frequency range at a spectral resolution of $\sim 35 \, {\rm kHz}$ with the ALMA Compact Array (ACA). The ACA data have similar angular resolution as the MUSTANG-2 data we present here, but  the nature of interferometry limits the scales recovered to $<90 \arcsec$ for the given projected antenna spacing in the 7-m array (the so-called `missing flux problem').  This is sufficient to compare the flux densities of the more or less discrete sources seen in the 3mm continuum maps, and to search for line contamination of the MUSTANG-2 data (\S~\ref{sec:linecontamination}).  The continuum fluxes are consistent to within a few percent.

\subsection{Instrumental Bandpasses}
\label{subsec:bandpasses}

With the exception of the {\it Herschel} data, all of the data
products used in this analysis are calibrated such that the
intensities represent average values across the instrument bandpass.
A spectral response curve $R(\nu)$ was obtained for
each instrument and its corresponding response-weighted center
frequency was calculated
\begin{equation}
\nu_0 = \frac{\int \, d\nu \, \nu \, R(\nu) }{\int \, d\nu \, R(\nu)}.
\end{equation}
\label{eq:freq}
To calculate the GBT telescope efficiency as
a function of frequency a $240 \, {\rm \mu m}$ (RMS) surface was
assumed \citep{hunter2011,frayer2019}. These response-weighted frequencies, shown in Table~\ref{tbl:freqs}, are used 
in evaluating spectral models (\S~\ref{subsec:modeling}). 
Compared to using nominal bandpass centers as representative frequencies, adopting these response-weighted 
frequencies will minimize differences between monochromatic flux
densities and band-averaged flux densities.  Section~\ref{sec:colorcorr} presents a
quantitative evaluation of the remaining effect. Table~\ref{tbl:freqs} also
presents the effective bandwidth for each instrument, which we define as:
\begin{equation}
\Delta \nu_{eff} = \frac{(\int \, d\nu \, R(\nu))^2}{\int \, d\nu \, R^2(\nu)}.
\end{equation}

\begin{table}
\centering
%\begin{center}
\begin{tabular} { |c|c|c| } \hline 
Instrument & $\nu_o$ [GHz] & $\Delta \nu_{eff}$ \\ \hline
GBT (1cm)   & $33.0$ & $14$ \\
GBT (MUSTANG-2) & $87.7$ & $28$ \\
GISMO & $150.6$ & $29$ \\
MAMBO & $254.9$ & $132$  \\
SCUBA-2 ($850 \, {\rm \mu m}$) & $354.0$ & $45$ \\
SCUBA-2 ($450 \, {\rm \mu m}$) & $668.1$ & $67$  \\
{\it Herschel} ($350 \, {\rm \mu m}$) & $871.9$ & $296$ \\
{\it Herschel} ($250 \, {\rm \mu m}$) & $1221.8$ & $420$ \\
{\it Herschel} ($160 \, {\rm \mu m}$) & $1910.7$ & $975$ \\ \hline
\end{tabular}
\caption{Response-weighted average frequencies and effective
  bandwidths for each instrument in this analysis.}
\label{tbl:freqs}
\end{table}

{\it Herschel} data are pipeline-calibrated to equivalent monochromatic flux
densities assuming a fiducial $\nu^{-1}$ spectrum.  We convert these
standard monochromatic flux densities $S_s (\nu_s)$ into
band-averaged flux densities $<S>$ using
\begin{equation}
<S> = S_s(\nu_s)  \times \frac{\int \, d\nu \, (\nu_s / \nu) \, R(\nu) }{\int \, d\nu \, R(\nu)}
\end{equation}
where $\nu_s$ is a chosen reference frequency, 
not in general equal
to the response-weighted  frequency given by Eq.~\ref{eq:freq}.
For {\it Herschel} the reference frequencies are those which
correspond to $\lambda = 350 \, {\rm \mu m}$, $250 \, {\rm \mu m}$,
and $160 \, {\rm \mu m}$.  These $\sim 1\%$ corrections were applied resulting in 
band-averaged flux densities.  

\section{Analysis}
\label{sec:analysis}

\subsection{SED Extraction}
\label{subsec:slices}

\begin{figure}
\centering
\includegraphics[width=\columnwidth, clip=true, trim=8.1mm 8mm 3.6mm 3.6mm]{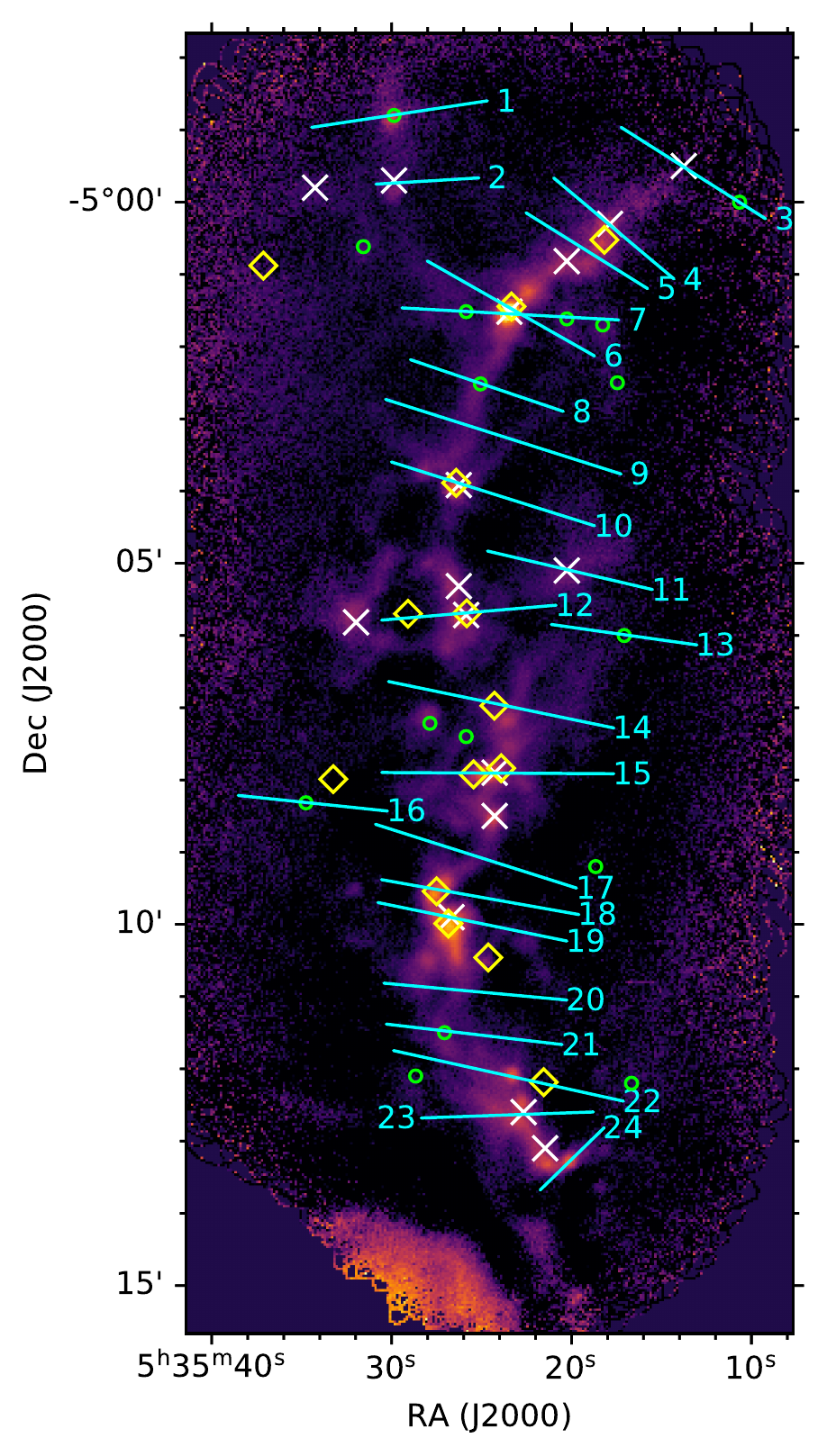}
\caption{MUSTANG-2 map, with the 24
  ``slices'' developed for detailed SED analysis shown in cyan. Green circles mark the locations of known starless cores in
  the field \citep{sadavoy2010}; yellow diamonds mark the locations
  where likely free-free emission has been identified from $8.3$~GHz VLA
  observations of R99; and white X's mark the locations
  of protostellar cores \citep{sadavoy2010}.}
\label{fig:snrWslices}
\end{figure}

We used the matched $25''$ resolution, multi-wavelength maps shown in Figure~\ref{fig:allDataWslices} to construct SEDs for a set of regions along the OMC 2/3 filament. 
Accurately extracting information from astronomical images made at different wavelengths, and with different telescopes and different data reduction algorithms, can be difficult due to the different systematics in the images. In order to cleanly focus on the spectrum of specific structures in
the maps, and to better account for variable local backgrounds and
zero levels in the maps, we have undertaken an analysis in
which we define a set of 24 lines or ``slices'' across structures of interest in the
maps.  The locations of these slices are shown in
Fig.~\ref{fig:snrWslices}; these slices were also used to define the pointings for
the 31 GHz photometry described in \S~\ref{sec:1cm}. In addition this figure  
shows the locations of protostars, starless cores, and
free-free emission from young stellar objects
\citep{sadavoy2010,Reipurth1999}.  The surface
brightness profile of each map along each slice is extracted, and we
remove a mean and slope from each profile at each wavelength. The mean
and slope are measured from the first and last 2\% of pixels along the
slice.  When we follow this procedure we find that the resulting
profiles, when renormalized to the peak intensity, track each other
remarkably well: the RMS of the individual-wavelength profiles about
the mean normalized profile is 9\% on average, with a highest
dispersion of 15\% (Slice 10) and a lowest dispersion of 5\% (Slice
6). This $9\%$ RMS scatter between the individual-wavelength profiles about their average is indicative of the maximum extent to which differences in the spatial filtering implicit in the maps is affecting the SEDs of the structures we are examining; it is an upper limit because astrophysical variations will also give rise to variations between the slice profiles at different wavelengths.
We take the peak value of the sky brightness across a slice to be the value of the Spectral Energy Distribution (SED) at that wavelength; in this manner the SED of the filament or object at the center of each slice is assembled, from 3~mm (MUSTANG-2) to $160 \, {\rm \mu m}$ ({\it Herschel}). The 1~cm SED points are determined from the pointed GBT nod observations,  re-normalized to the same surface brightness units.
Calibration uncertainties dominate the spectral analysis: we assume a 10\% calibration uncertainty associated with each SED point. For the 1~cm data the measurement error is added in quadrature.

%{\bf To Do: try break at 130 GHz and 60 GHz, compare goodness of fit.  try a physically motivated eg FDS99 double MBB model. Again, goodness of fit.}
%  References for flattening beta at long lambda: Planck intermediate results XIV; Finkbeiner 99; Galliano+ 06; Paladini+07; 
% Planck XVII 2011; Planck XIX 2011; TLS = Meny+07; magnetic dipole = Draine&Hensley 13; carbon dust Jones+13

\begin{figure*}
\centering
\gridline{\fig{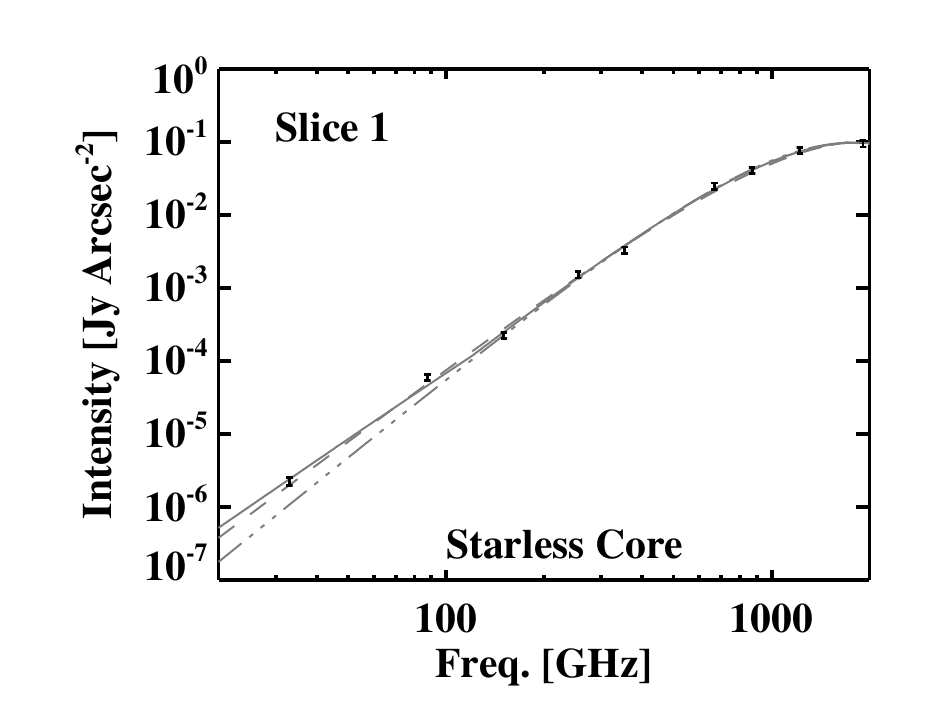}{\columnwidth}{}
\fig{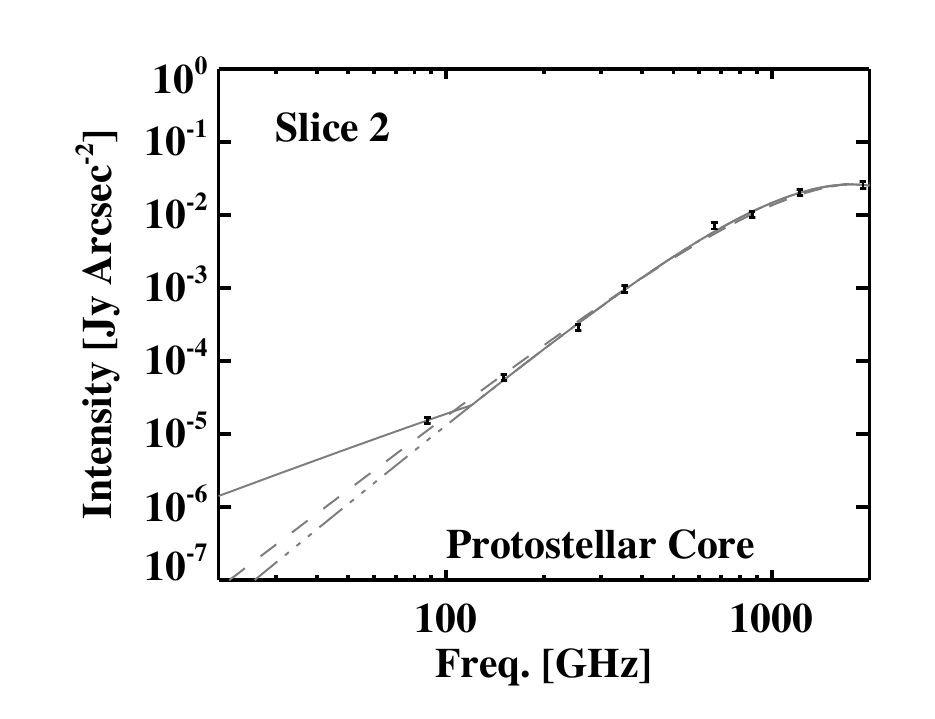}{\columnwidth}{}}
\gridline{\fig{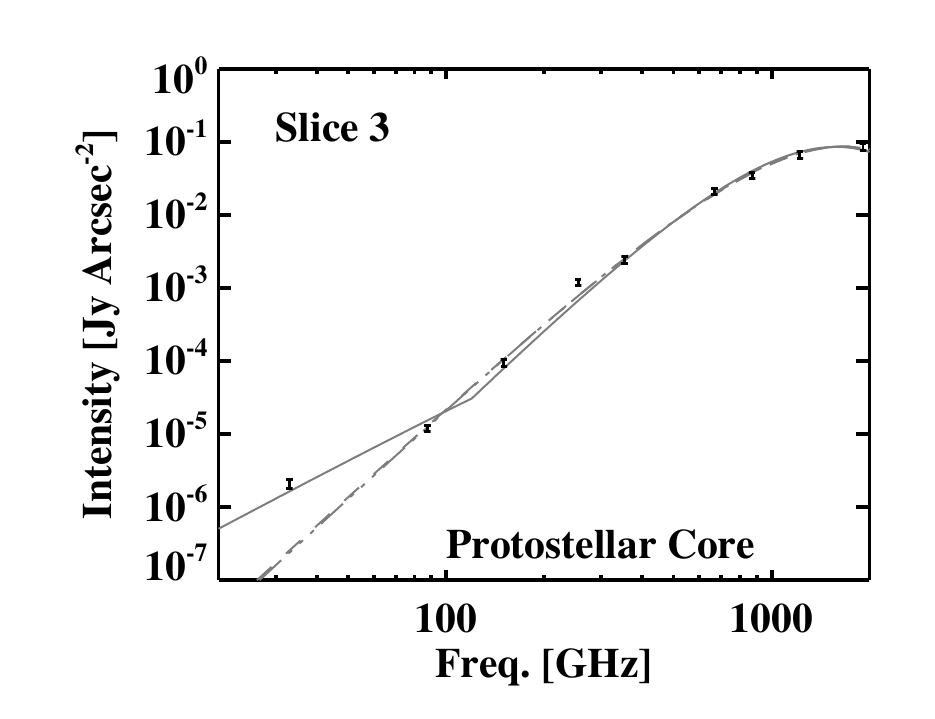}{\columnwidth}{}
          \fig{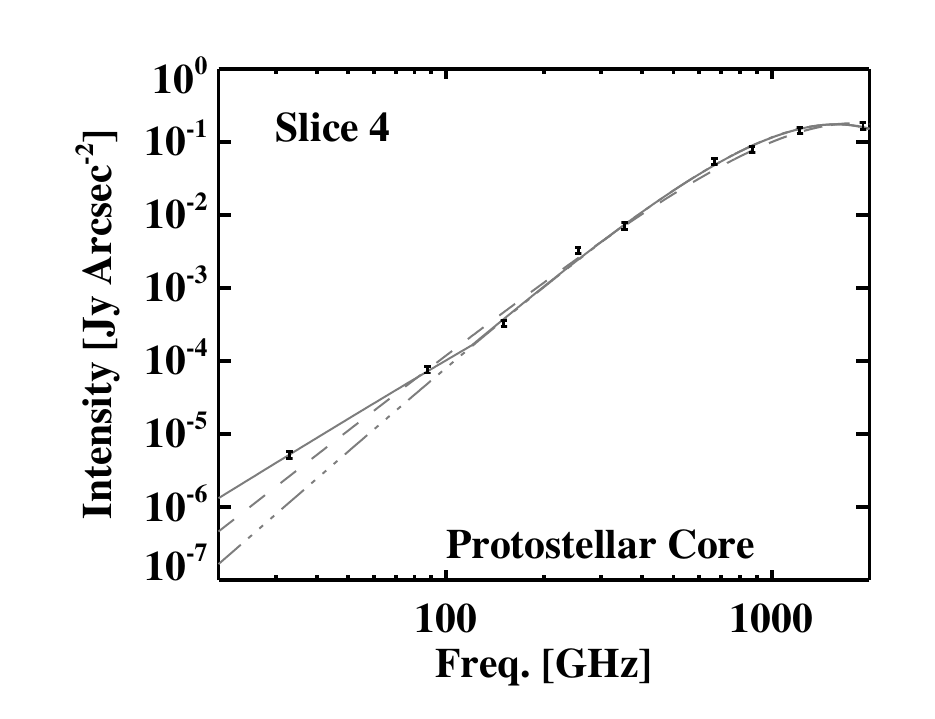}{\columnwidth}{}}
\gridline{\fig{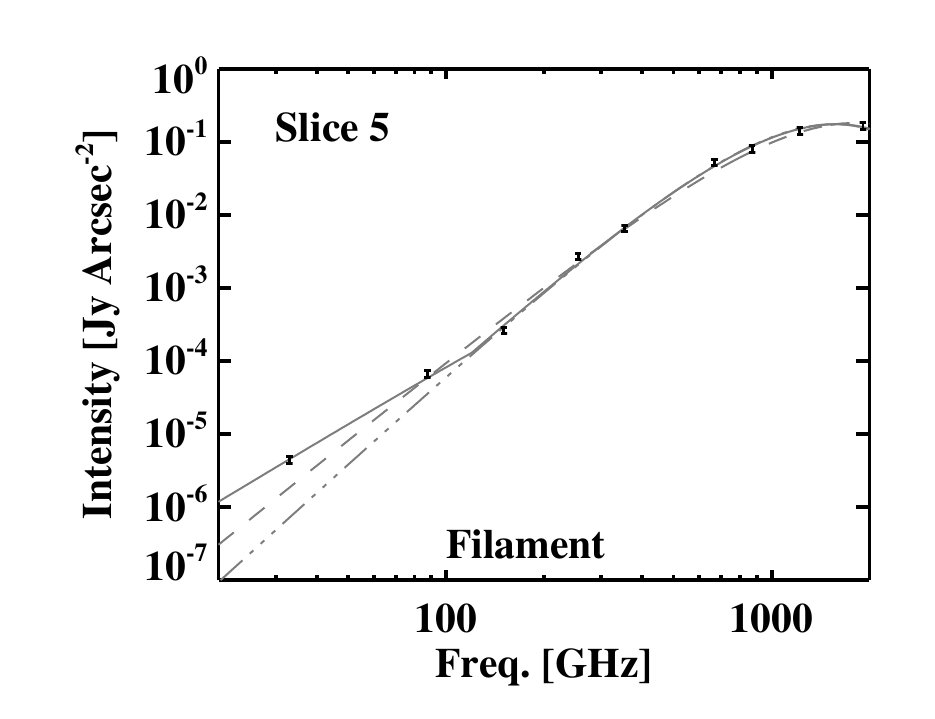}{\columnwidth}{}
          \fig{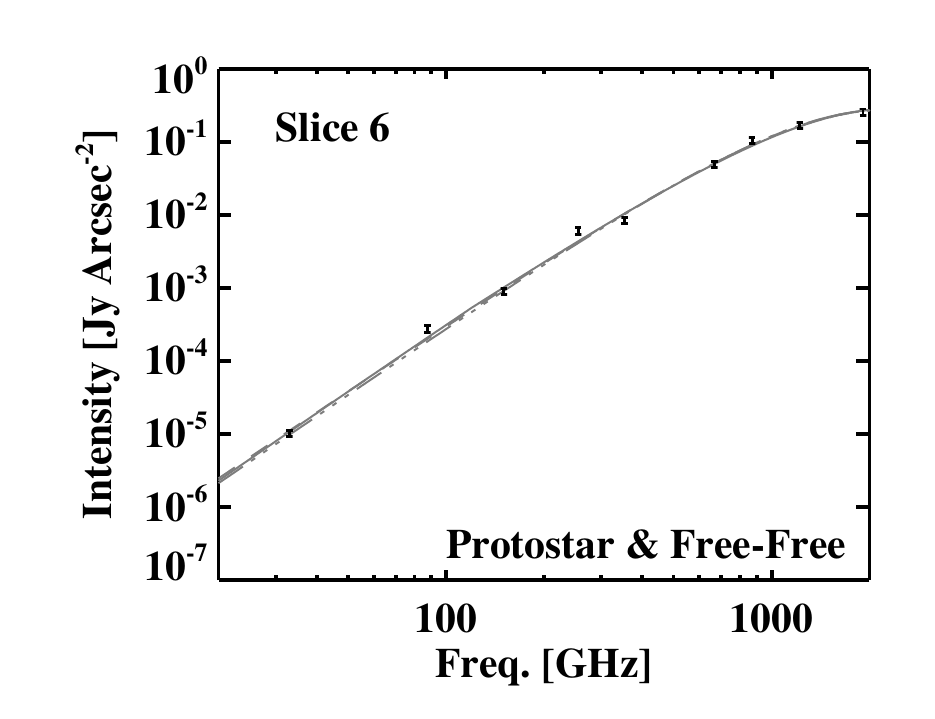}{\columnwidth}{}}
\caption{Spectral Energy Distribution for slices 1-6.  The solid line is the broken modified black body fit to the data; the dashed line is a standard modified black body fit to the data; and the dash-triple-dot line is a standard modified black body fit to the $\lambda \leq 2 \, {\rm mm}$ data only. }
\label{fig:sedplots1}%
\end{figure*}

\begin{figure*}
\centering
\gridline{\fig{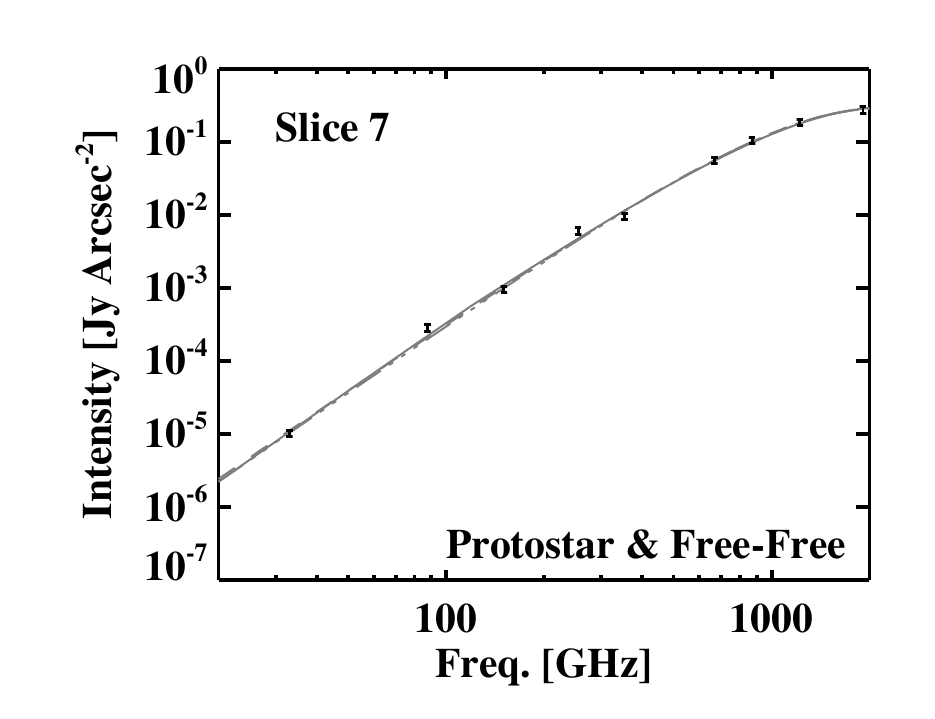}{\columnwidth}{}
          \fig{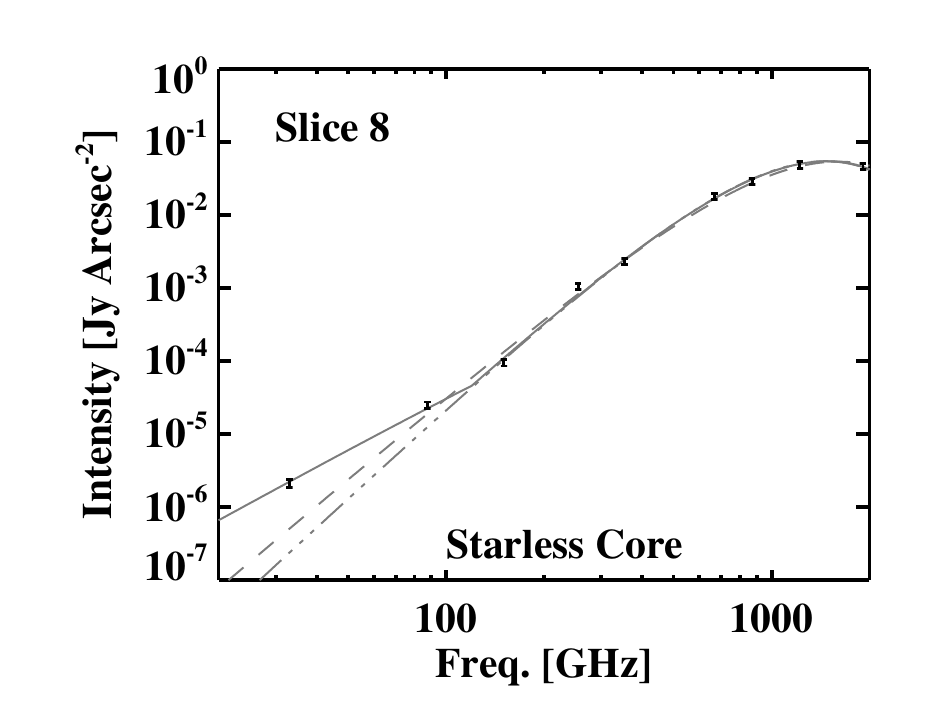}{\columnwidth}{}}
\gridline{\fig{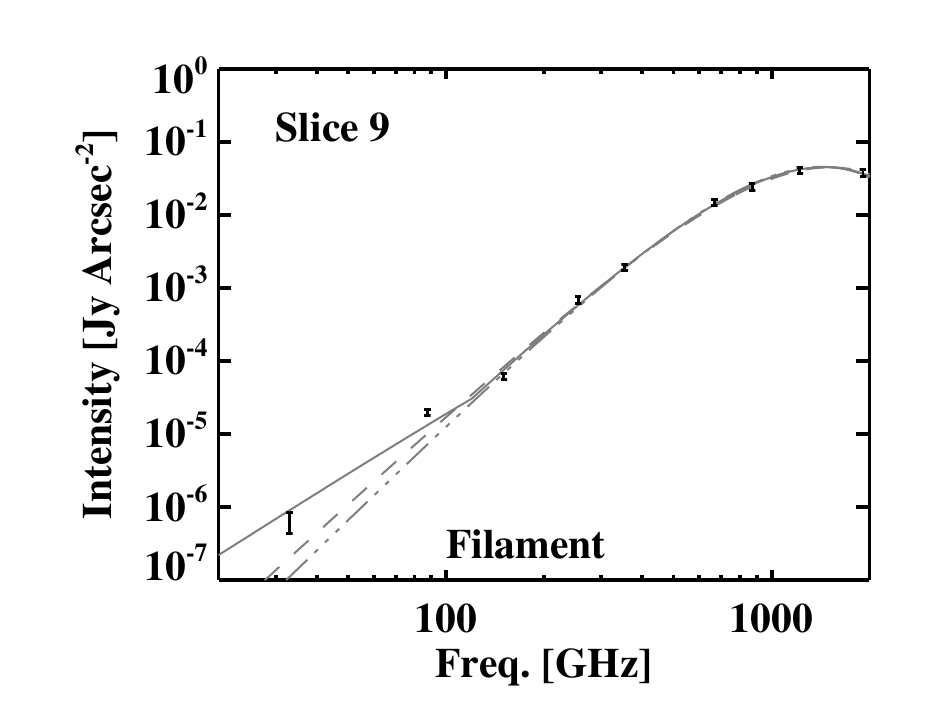}{\columnwidth}{}
          \fig{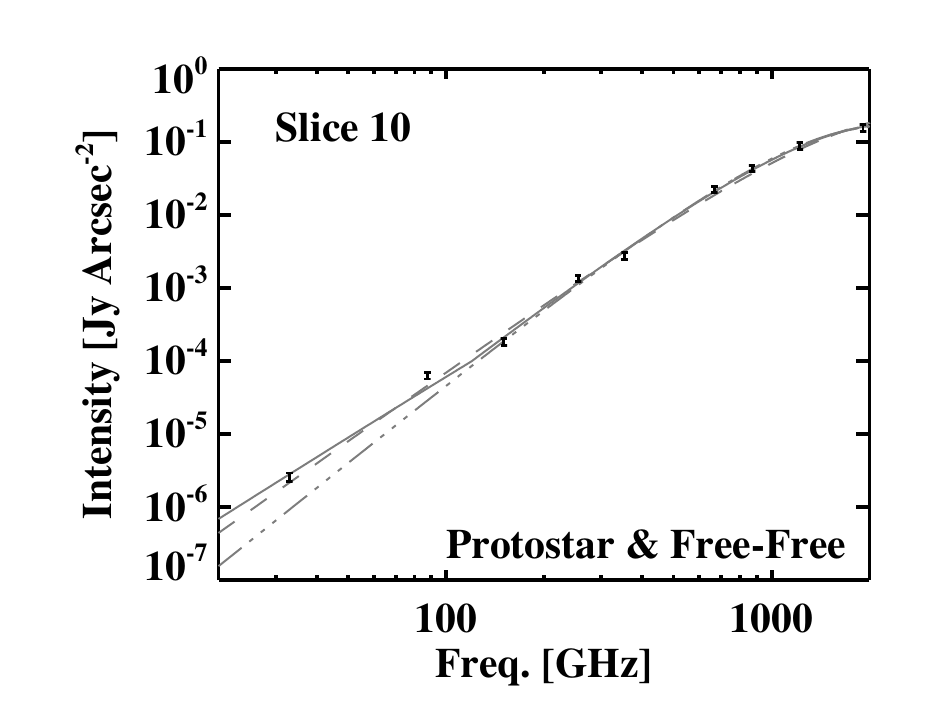}{\columnwidth}{}}
\gridline{\fig{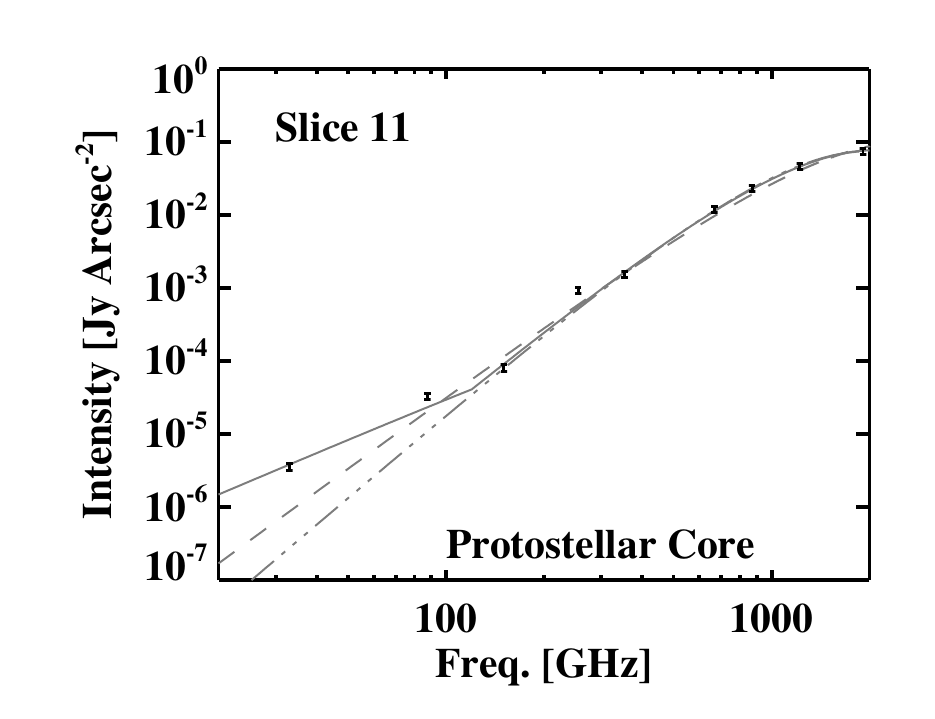}{\columnwidth}{}
          \fig{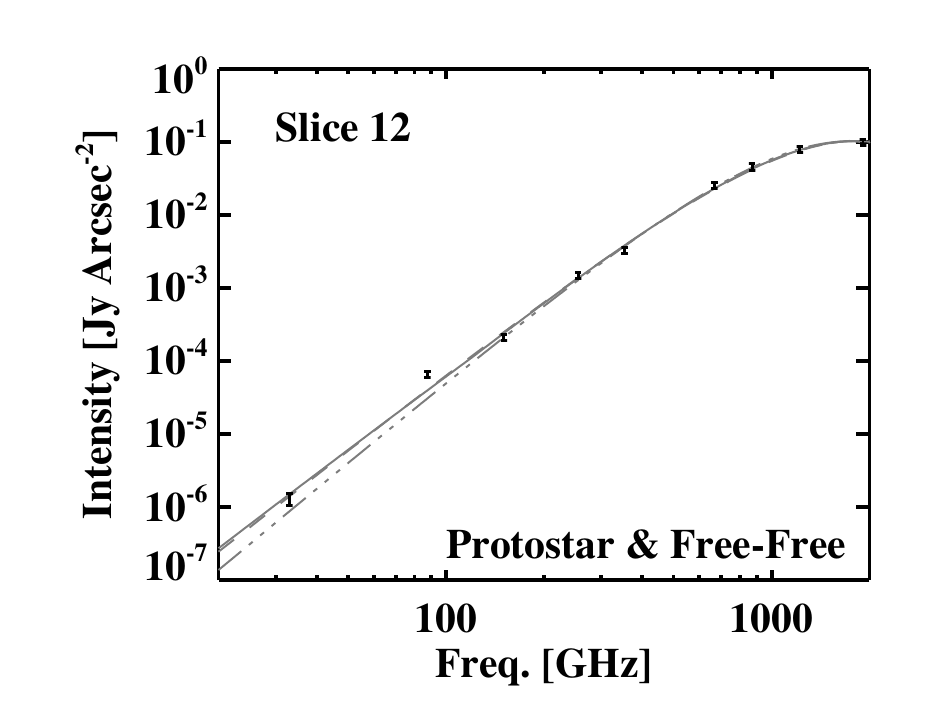}{\columnwidth}{}}
\caption{Spectral Energy Distribution for slices 7-12. Lines are as in Figure~\ref{fig:sedplots1}.}
\label{fig:sedplots2}%
\end{figure*}

\begin{figure*}
\centering
\gridline{\fig{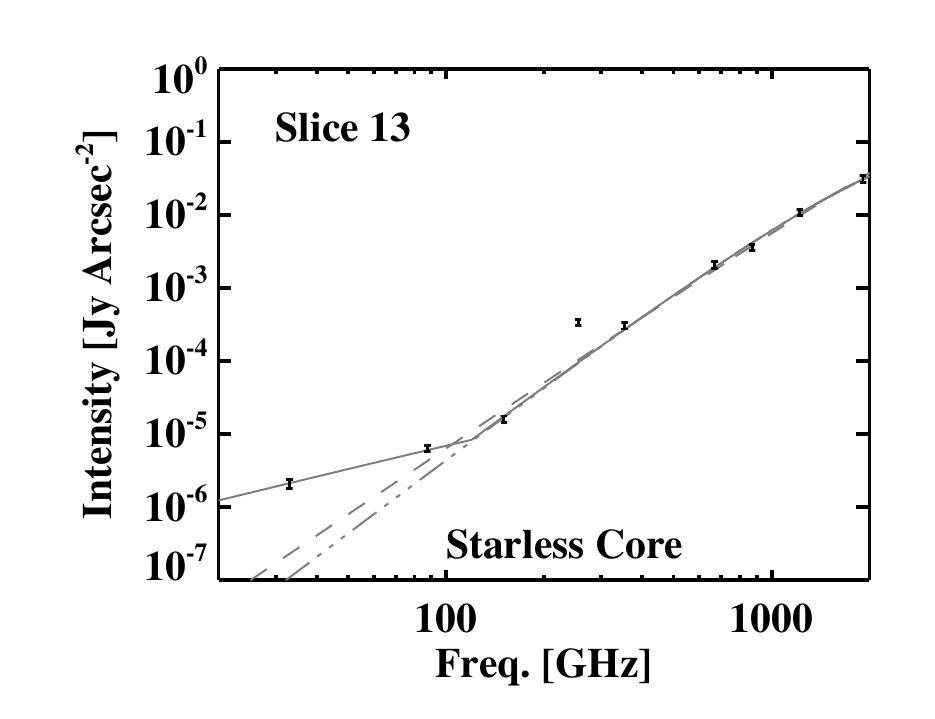}{\columnwidth}{}
          \fig{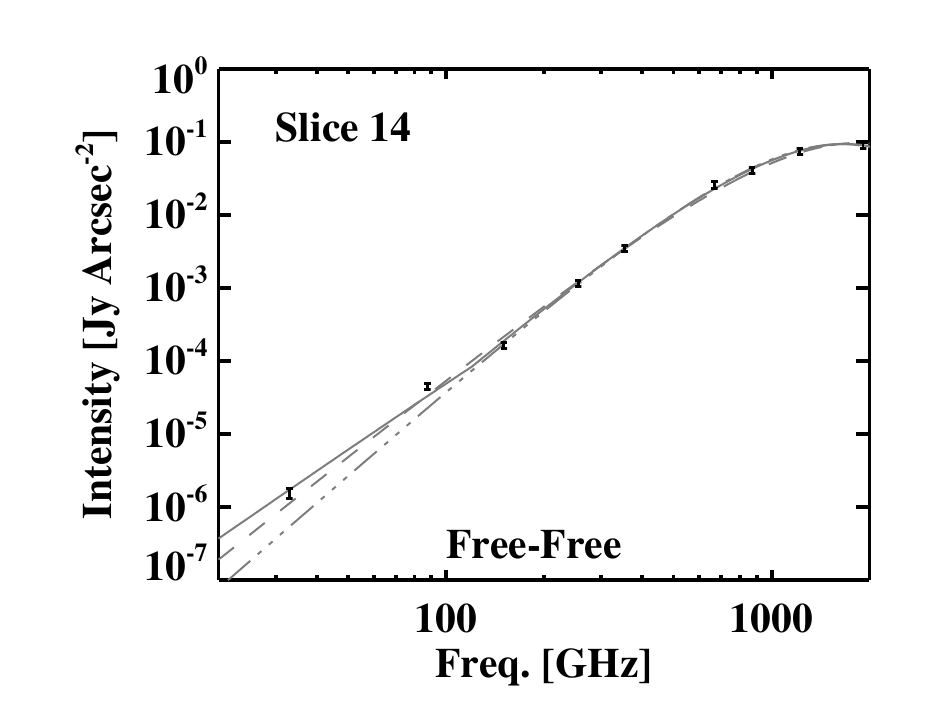}{\columnwidth}{}}
\gridline{\fig{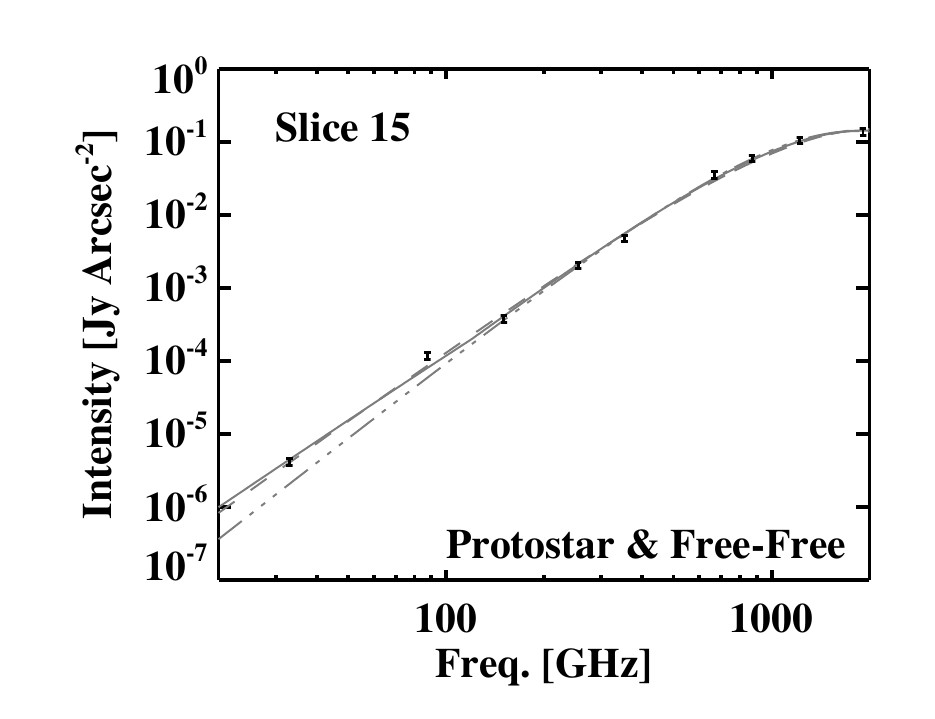}{\columnwidth}{}
          \fig{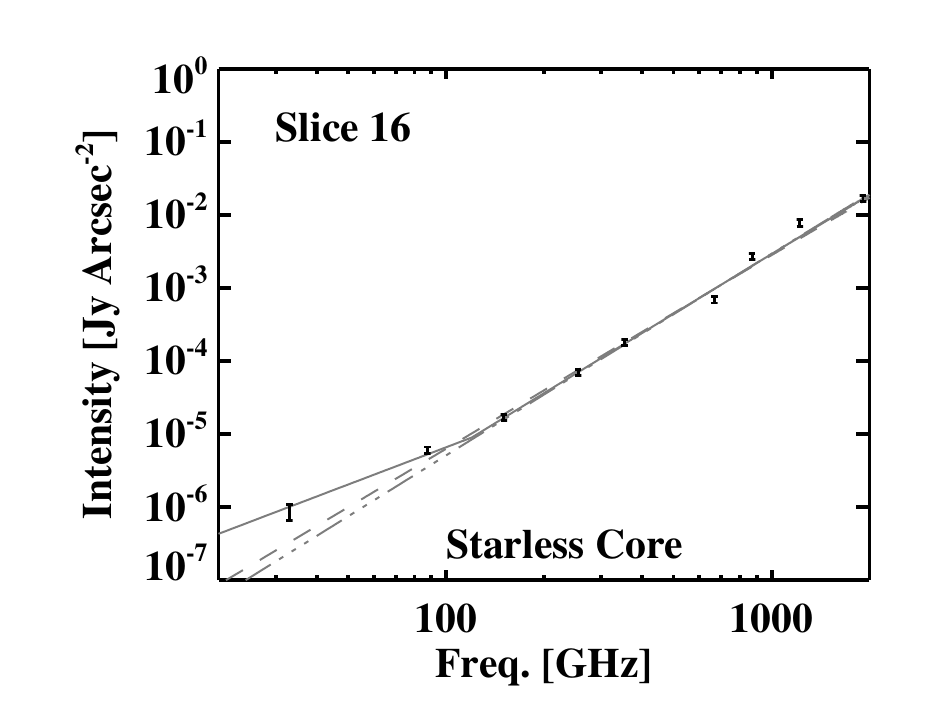}{\columnwidth}{}}
\gridline{\fig{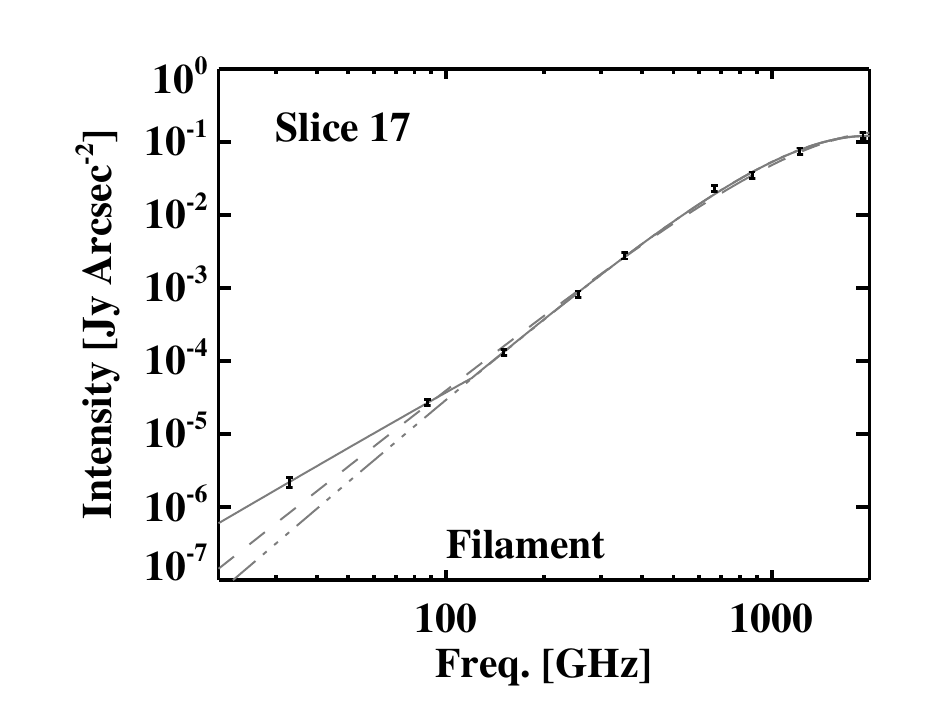}{\columnwidth}{}
          \fig{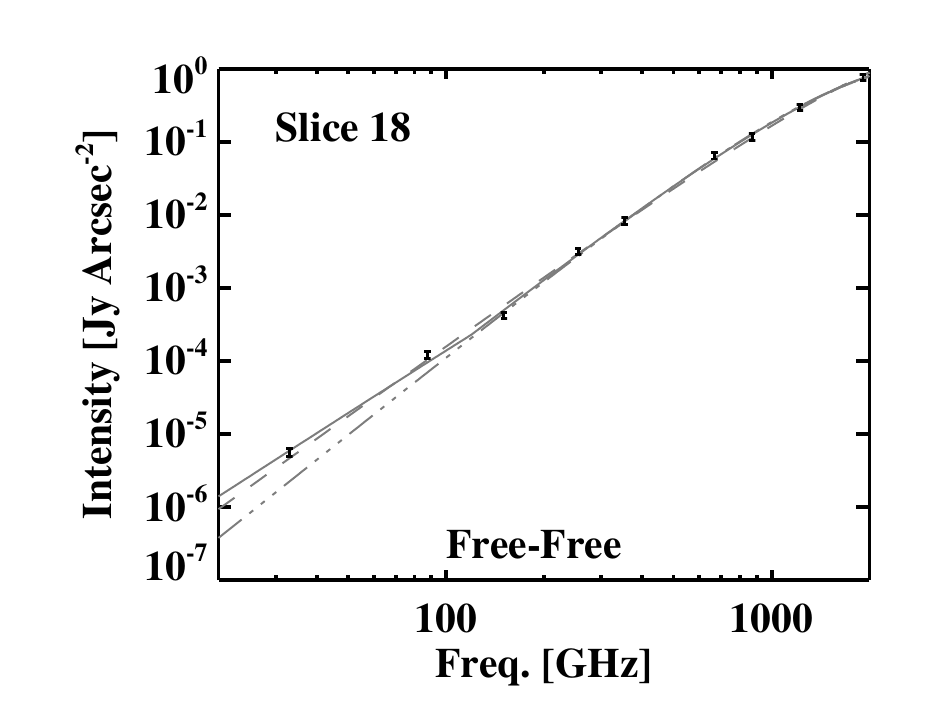}{\columnwidth}{}}
\caption{Spectral Energy Distribution for slices 13-18. Lines are as in Figure~\ref{fig:sedplots1}.}
\label{fig:sedplots3}%
\end{figure*}

\begin{figure*}
\centering
\gridline{\fig{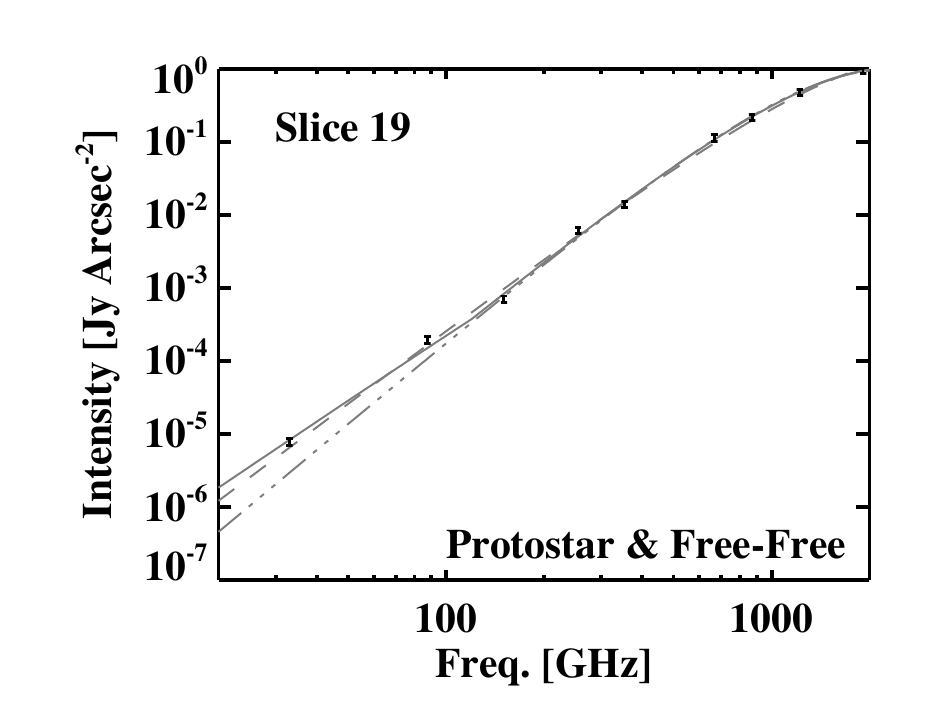}{\columnwidth}{}
          \fig{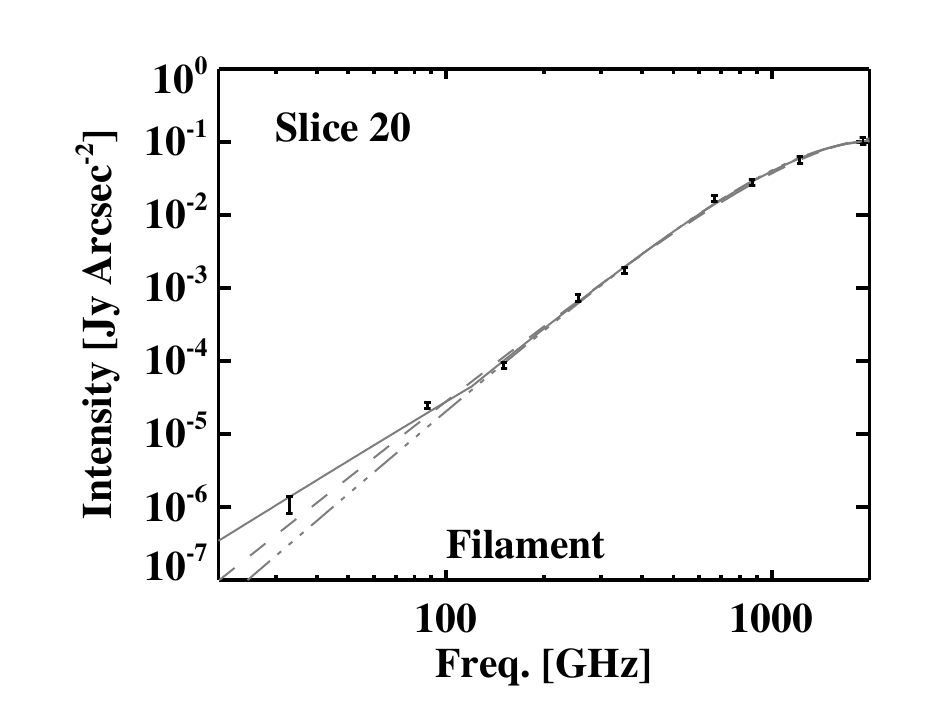}{\columnwidth}{}}
\gridline{\fig{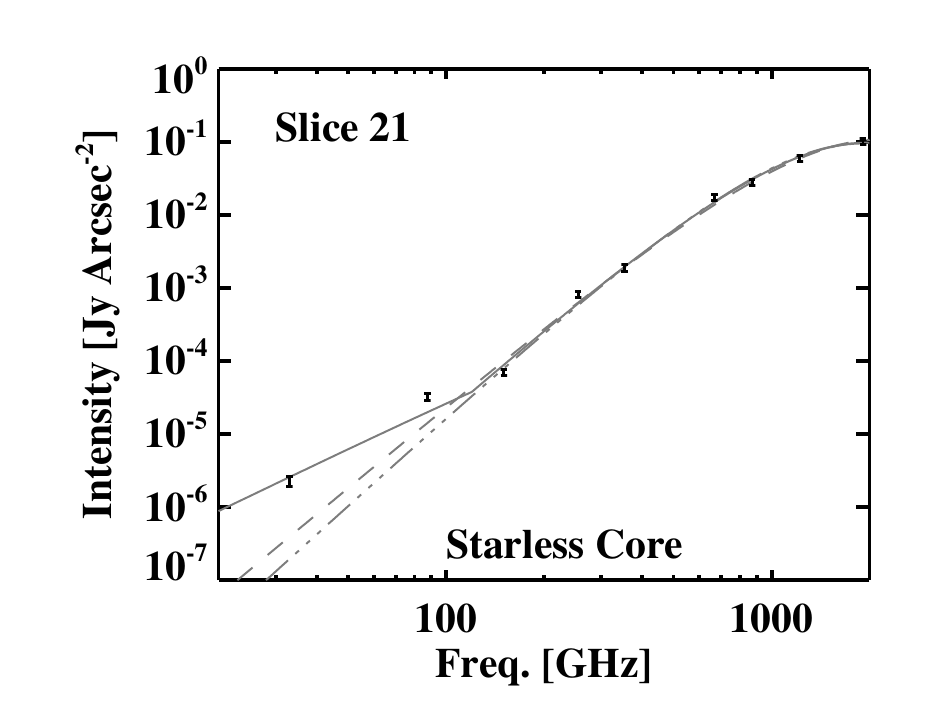}{\columnwidth}{}
          \fig{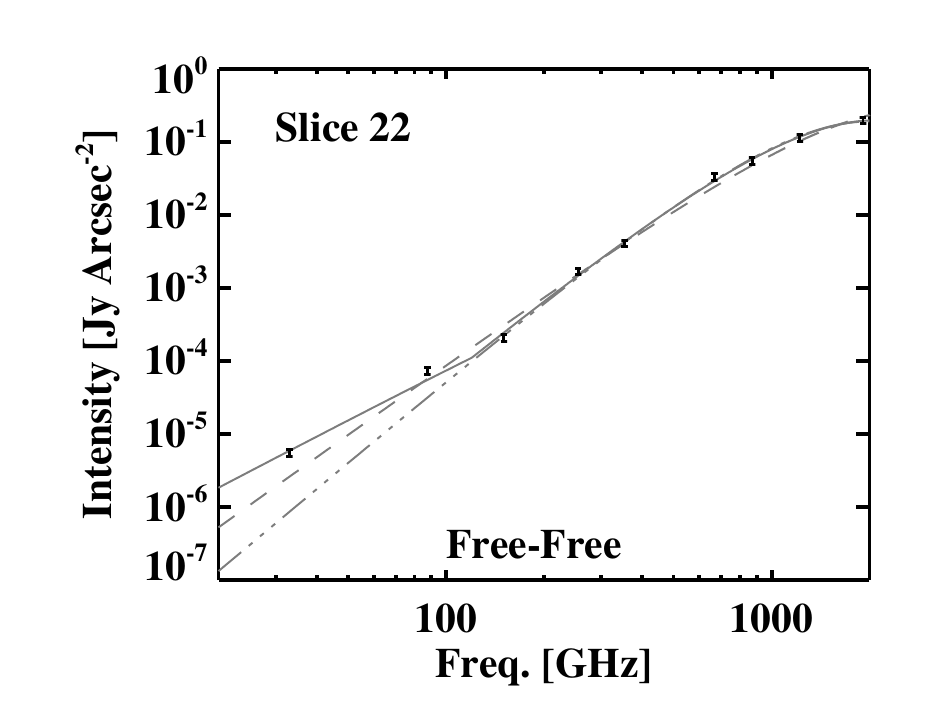}{\columnwidth}{}}
\gridline{\fig{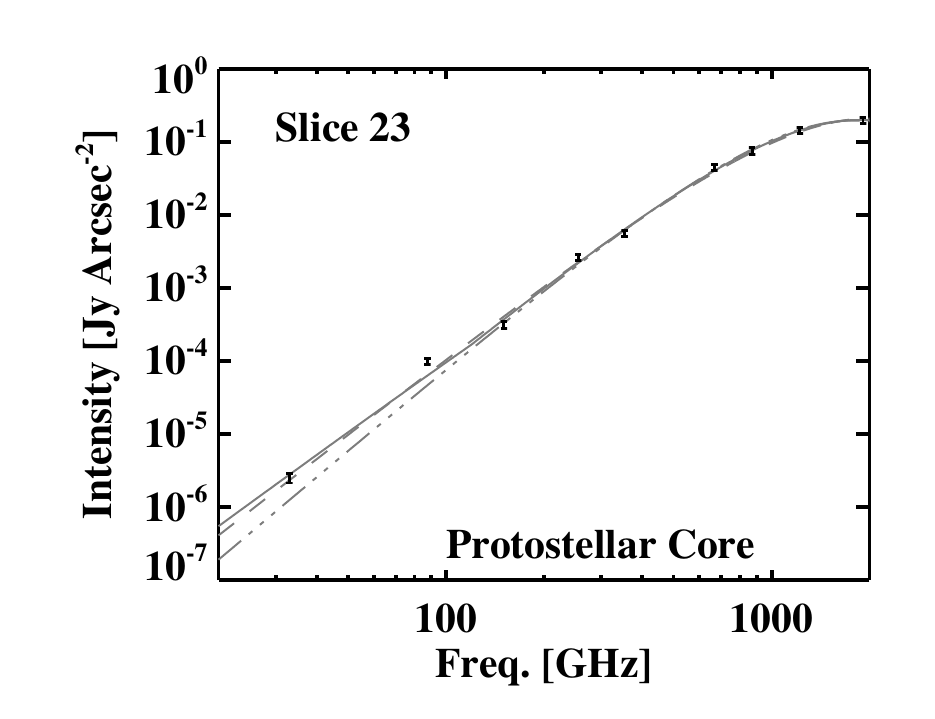}{\columnwidth}{}
  \fig{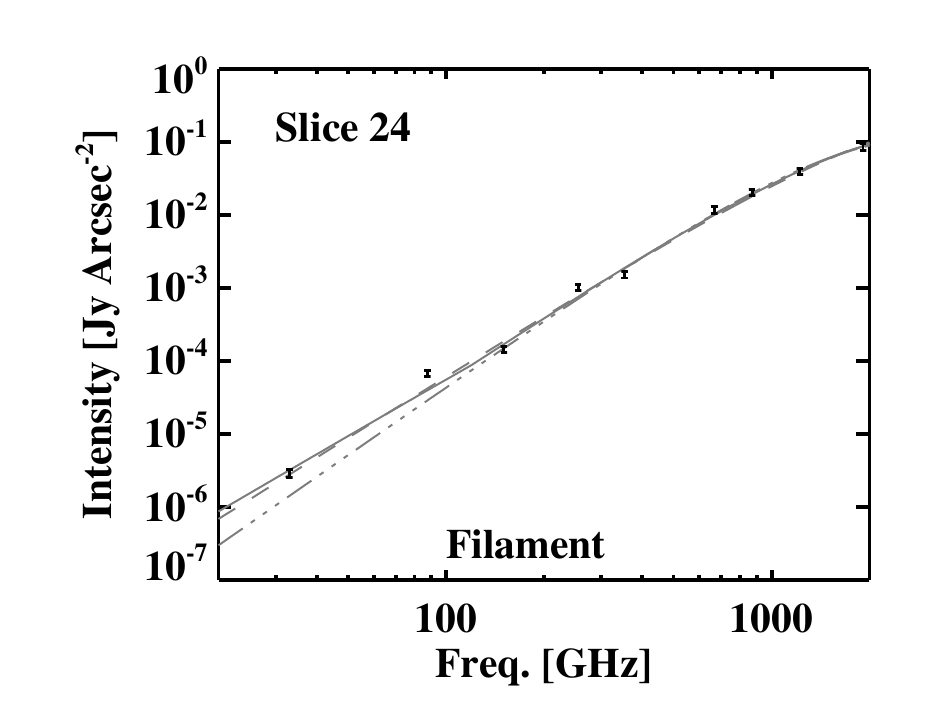}{\columnwidth}{}
}
\caption{Spectral Energy Distribution for slices 19-24. Lines are as in Figure~\ref{fig:sedplots1}.}
\label{fig:sedplots4}%
\end{figure*}

The SEDs for the 24 regions of OMC 2/3 studied here are shown in Figures~\ref{fig:sedplots1} through \ref{fig:sedplots4}.  In general, and as discussed in \S~\ref{subsec:modeling}, the $\lambda > 2 \, {\rm mm}$ data are seen to be higher than expected based on extrapolations from shorter wavelength.
Salient characteristics
of some of the individual regions are as follows.
\begin{itemize}
    \item Slice 2: GBT 31 GHz data rejected due to contamination in the reference beam.
    \item Slice 6 \& 7: redundant measurements of the brightest object in the 3~mm map, MMS6. The peak brightness at 3~mm was $90 \, {\rm mJy}$ per $10''$ beam.  
    \item Slices 11 \& 13: the 1~mm data point was anomalously high. Since these are isolated anomalies, and since
    the 1~mm map is clearly discrepant with the morphologies evident at other wavelengths in this region, the 1~mm data points were excluded from analysis.
\end{itemize}

\subsection{Modeling}
\label{subsec:modeling}

The Modified Blackbody (MBB) is the spectrum generated by 
optically thin thermal radiation from dust grains at a temperature $T_d$  and having a power law opacity proportional to $\nu^{\beta}$: 
\begin{equation}
    I(\nu) \propto \frac{\nu^{(3 +\beta)}}{exp(h\nu/k T_d)-1}.
\end{equation}
Given the elevated long-wavelength emission we seek to quantify, we have studied two different
selections of our data with two different variants of the MBB. The four cases are as follows:
\begin{itemize}
    \item MBB2mm: we fit the $\lambda \le 2 \, {\rm mm}$ data to a standard Modified Blackbody.
    \item MBB2mm\_ext: uses the models resulting from MBB2mm, but evaluating goodness of fit with respect to the entire SED, {\it i.e.} the fits are extrapolated to longer wavelength.
    \item MBBall: all SED points for a given target are jointly fit to a single MBB.
    \item MBBall\_2beta: same as MBBall, but a spectral break is introduced at $120 \, {\rm GHz}$ with an independent emissivity index $\beta'$ to model long wavelength emission.
\end{itemize}
The $\chi^2$ per degree of freedom and number of degrees of freedom $n$ for each of these analyses are given in Table~\ref{tbl:chi2}, and each analysis is discussed in more detail in the following sections.   For purposes of fitting and evaluating $\chi^2$, the MBB is evaluated at each instrument's response-weighted frequency $\nu_0$ (\S~\ref{subsec:bandpasses}). The effect of residual band-averaging effects (or color corrections) on the model predictions is quantified in \S~\ref{sec:colorcorr}.

\subsubsection{Fitting $\lambda \leq 2 \, {\rm mm}$ (MBB2mm, MBB2mm\_ext)}

We first attempt a standard Modified Blackbody (MBB) fit, as done
before for very similar data in S16.  This is the analysis we call
MBB2mm, and we find it gives reasonable fits to the 2~mm and shorter
wavelength data, with a median over the 24 ``slices'' of $\chi_{n}^2 =
\chi^2/n = 1.65$ for $n = 4$ degrees of freedom, a minimum $\chi_{n}^2
= 0.11$ and a maximum $\chi_{n}^2 = 9.11$.

When we extrapolate these fits to the entire measured SED, however, we
find considerably poorer consistency. This is the case we call
MBB2mm\_ext. It has a median $ \chi_n^2$ of $9.48$ for $n = 6$ degrees
of freedom, a minimum $\chi_{n}^2 = 3.09$ and a maximum $\chi_{n}^2 =
22.26$.  The poor fits are due to the 3~mm and 1~cm data points
systematically lying above the extrapolated fits to the short
wavelength data: 23 of 24 3~mm SED points lie above the short
wavelength MBB fit, with a median excess of $5.0 \sigma$; 22 of these
23 have a $3\sigma$ or greater excess.  Similarly, 22 of 23 of the
1~cm data points lie above the extrapolated MBB, also with a typical
magnitude of $5.0\sigma$, and the single point below the fit is not
statistically significant ($0.1 \sigma$).  The average residual
brightness is $(1.96 \pm 0.35) \times 10^{-6} \, {\rm Jy \,
  Arcsec^{-2}}$ at 1~cm and $(3.28 \pm 0.64) \times 10^{-5} \, {\rm Jy
  \, Arcsec^{-2}}$ at $3.3 {\rm mm}$.  In fractional terms the 1~cm
data points themselves are $5.19 \pm 1.03$ times as bright as the
extrapolated MBB, and the $3.3 \, {\rm mm}$ data are $2.01 \pm 0.15$
times as bright as the model extrapolation.

% RECOUNT RESIDUALS IE ABOVE BELOW STATS I HAD 24 INSTEAD OF 23 AND 22 INSTEAD OF 23.

\subsubsection{Fitting all the data (MBBall, MBBall\_2beta)}

Including the long wavelength points in the MBB fit still yields poor
$\chi^2_n$ values, with a median $\chi_{n}^2 = 5.23$ (minimum =
$2.76$, maximum = $17.44$) for $n = 6$ and the long wavelength data
systematically above the model.  This is the case we call MBBall. In
this case 19 of 23 1~cm SED points lie above the fit (typical SNR
$=1.9\sigma$) and 23 of 24 3~mm SED points lie above the fit (typical
SNR $2.7\sigma$).

To better accommodate the long wavelength data we allow a different
power law $\beta'$ at $\nu < 120 \, {\rm GHz}$ ($\lambda > 2.5 \, {\rm
  mm}$; we refer to this case as MBBall\_2beta. The break frequency
was chosen to lie midway between the GISMO and MUSTANG-2 bandpass
centers. For discussion of a physical motivation of this
generalization, see \S~\ref{subsec:dust}. We find a considerably
improved fit, with a median $\chi_{n}^2 = 2.98$ for $n = 5$ (minimum
$=0.84$, maximum $=7.79$).  The residuals to the MBBall\_2beta fit
still show systematic trends, with 22 of the 23 1~cm SED points
falling {\it below} the fit and 23 of the 24 3~mm SED points still
falling above it.  This suggests that the spectral shape at $\lambda >
2 \, {\rm mm}$ may not be well-represented by a pure, single power
law. This fact is also reflected in the lower $\chi^2_n$ for
the single blackbody fit to the short-wavelength data (MBB2mm: $1.65$)
compared to the double-beta blackbody fit to {\it all} the data
(MBBall\_2beta: $2.98$).  The limited limited spectral coverage of our
dataset in this regime however would not support introducing more
parameters into the analysis, and no single, plausible physical model
provides a better description of the long wavelength residuals
(\S~\ref{sec:interpretation}). Nevertheless,  of the models we considered the double-beta modified blackbody model clearly provides the best description of the dataset
as a whole.

%{\bf the alldata BBB fit: are somewhat driven by the Ka data since error bars in absolute terms are small. only 3 Ka points are %*above* the fit.  19 of 3~mm points are above the fit so shape is clearly not pure power law}

Table~\ref{tbl:chi2} shows $\chi^2_{n}$ and $n$ values for all four of the modeling approaches described above. The $\chi^2_{n}$ of MBBall\_2beta is lower than that in the MBBall analysis for 19 of 24 regions.  The statistical significance of the reduction in $\chi^2$ achieved by allowing an extra parameter ($\beta'$) to be free in the model fit can be evaluated by the F-test. The F statistic is computed as
\begin{equation}
    F = \frac{\Delta \chi^2}{\chi^2_{n,u}}
\end{equation}
where $\Delta \chi^2$ is the change in $\chi^2$ which has resulted from adding the additional parameter(s), and $\chi^2_{n,u}$ is the reduced $\chi^2$ for the ``unrestricted'' model fit, in this case, the one which also allows $\beta'$ to be a free parameter. For eight individual regions the reduction in $\chi^2_{n}$ is significant at the 95\% level by this test; these regions are indicated by bold entries in Table~\ref{tbl:chi2}.  The physical environments of these eight regions are varied: three are starless cores, two are protostellar cores, one has a known free-free source, and two sample filamentary dust emission. In no case is the simple MBB significantly preferred over MBBall\_2beta.

\begin{table*}
%\centering
\begin{center}
\begin{tabular} { |c|c|c|c|c| } \hline 
 & \multicolumn{4}{c}{$\chi^2/n \, \, \, (n)$}  \\ \hline
 & MBB2mm & MBB2mm\_ext & MBBall & MBBall\_2beta \\
Slice  & ($\lambda \le 2 \, {\rm mm}$) & (All Data, extrap.) & (All Data, fit) & (All Data) \\  \hline
       1   &   $1.12$       (4) &      $6.24$       (6) &      $2.76$       (6) &      $2.27$       (5) \\  
       2   &   $1.50$       (4) &      $5.53$       (5) &      $3.84$       (5) &      $1.50$       (4) \\  
       3   &   $3.90$       (4) &      $9.48$       (6) &      $9.44$       (6) &      $7.20$       (5) \\  
       4   &   $2.89$       (4) &      $12.80$       (6) &      $8.22$       (6) &      ${\bf 2.39}$       (5) \\  
       5   &   $2.48$       (4) &      $14.97$       (6) &      $10.29$       (6) &      ${\bf 2.39}$       (5) \\  
       6   &   $4.20$       (4) &      $4.68$       (6) &      $4.25$       (6) &      $4.94$       (5) \\  
       7   &   $2.42$       (4) &      $3.09$       (6) &      $2.91$       (6) &      $3.28$       (5) \\  
       8   &   $2.51$       (4) &      $12.99$       (6) &      $10.26$       (6) &      ${\bf 2.31}$       (5) \\  
       9   &   $1.68$       (4) &      $8.93$       (6) &      $7.33$       (6) &      $4.20$       (5) \\  
      10   &   $1.61$       (4) &      $10.41$       (6) &      $5.23$       (6) &      $4.00$       (5) \\  
      11   &  $0.11$       (3) &      $22.26$       (5) &      $17.44$       (5) &      ${\bf 2.98}$       (4) \\  
      12   &   $1.13$       (4) &      $5.98$       (6) &      $4.08$       (6) &      $4.79$       (5) \\  
      13   &   $1.52$       (3) &      $16.64$       (5) &      $13.89$       (5) &      ${\bf 1.31}$       (4) \\  
      14   &  $0.50$       (4) &      $7.15$       (6) &      $3.85$       (6) &      $2.23$       (5) \\  
      15   &  $0.81$       (4) &      $8.19$       (6) &      $3.21$       (6) &      $3.20$       (5) \\  
      16   &   $9.11$       (4) &      $10.44$       (6) &      $9.25$       (6) &      $7.79$       (5) \\  
      17   &   $1.03$       (4) &      $6.75$       (6) &      $4.60$       (6) &     ${\bf 0.84}$       (5) \\  
      18   &   $1.10$       (4) &      $8.69$       (6) &      $3.66$       (6) &      $2.09$       (5) \\  
      19   &   $1.46$       (4) &      $10.22$       (6) &      $4.35$       (6) &      $2.53$       (5) \\  
      20   &   $1.72$       (4) &      $6.57$       (6) &      $4.59$       (6) &      $2.73$       (5) \\  
      21   &   $3.65$       (4) &      $16.54$       (6) &      $14.63$       (6) &      ${\bf 6.41}$       (5) \\  
      22   &   $1.41$       (4) &      $16.09$       (6) &      $10.75$       (6) &      ${\bf 2.76}$       (5) \\  
      23   &   $1.65$       (4) &      $7.68$       (6) &      $4.43$       (6) &      $4.48$       (5) \\  
      24   &   $3.81$       (4) &      $10.39$       (6) &      $6.58$       (6) &      $7.15$       (5) \\  \hline
      Median & $1.65$ & $9.48$ & $5.23$ & $2.98$ \\
      Min. & $0.11$  & $3.09$  & $2.76$ & $0.84$ \\
      Max. & $9.11$  & $22.26$ & $17.44$ & $7.79$ \\ \hline
\end{tabular}
\caption{$\chi^2_{n} = \chi^2 / n$ for the four fits described in the text (\S~\ref{sec:analysis}): the short wavelength MBB fit (MBB2mm); the short wavelength MBB fit evaluated relative to all data (MBB2mm\_ext); the MBB fit to all data (MBBall);  and a ``Broken Modified Blackbody'' fit to all data (MBBall\_2beta).  The number of degrees of freedom $n$ is given for each in parentheses. The eight regions for which the broken black body fit (MBBall\_2beta) has a $\chi_{n}^2$  lower than the all-data MBB $\chi_{n}^2$ at 95\% confidence are indicated by bold values in the last column. }
\label{tbl:chi2}
\end{center}
\end{table*}

Table~\ref{tbl:fitparams} shows the fitted parameter values from the
MBBall\_2beta analysis for each region studied, and
Table~\ref{tbl:avgparams} summarizes the typical parameters for each
of the types of objects we studied.  The values of $\beta$ and $T_d$
obtained are generally consistent with those obtained in this region
by measurements of ammonia transitions \citep{Friesen2017}, as well as
by the similar mm/sub-mm continuum study of this region by S16.  The
dust temperatures are coldest ($15.2 \pm 0.5 \, {\rm K}$) in the
filament, and progressively warmer in starless cores, cores with
protostars, and finally warmest in regions with detected free-free
sources ($20.4 \pm 0.6 \, {\rm K}$).  The long wavelength $\beta'$
averages $0.74 \pm 0.03$, though it is strikingly flatter for starless
cores than in other regions ($\beta' = 0.21 \pm 0.06$).

\vspace{0.25in}

\begin{table*}
%\centering
\begin{center}
\begin{tabular} { |c|c|c|c|c|} 
\hline 
Slice Number	&	Description	&	$\beta$			&	$\beta'$			&	$T_d$			\\ \hline
1	&	Starless Core	                         &  $1.52 \pm  0.10$ &  $1.09 \pm  0.12$ & $ 19.44 \pm  1.44$ \\
2	&	Protostellar Core	                  &  $1.67 \pm  0.12$ & -$0.32 \pm  0.45$ & $ 18.33 \pm  1.39$ \\
3	&	Protostellar Core	                  &  $2.43 \pm  0.10$ &  $0.39 \pm  0.16$ & $ 14.04 \pm  0.74$ \\
4	&	Protostellar Core	                  &  $1.89 \pm  0.10$ &  $0.78 \pm  0.11$ & $ 15.67 \pm  0.91$ \\
5	&	Filament	                          &  $2.03 \pm  0.10$ &  $0.71 \pm  0.11$ & $ 15.16 \pm  0.86$ \\
6	&	Protostellar Core, Free-Free Emission     &  $0.94 \pm  0.10$ &  $1.09 \pm  0.10$ & $ 28.25 \pm  3.31$ \\
7	&	Protostellar Core, Free-Free Emission     &  $0.97 \pm  0.10$ &  $1.15 \pm  0.10$ & $ 27.09 \pm  3.00$ \\
8	&	Starless Core	                          &  $2.10 \pm  0.10$ &  $0.47 \pm  0.12$ & $ 14.01 \pm  0.71$ \\
9	&	Filament	                          &  $2.27 \pm  0.10$ &  $0.86 \pm  0.19$ & $ 13.34 \pm  0.65$ \\
10	&	Protostellar Core, Free-Free Emission     &  $1.44 \pm  0.10$ &  $0.83 \pm  0.11$ & $ 25.88 \pm  2.75$ \\
11	&	Protostellar Core	                  &  $1.68 \pm  0.10$ & -$0.09 \pm  0.11$ & $ 21.54 \pm  1.83$ \\
12	&	Protostellar Core, Free-Free Emission     &  $1.57 \pm  0.10$ &  $1.43 \pm  0.14$ & $ 19.19 \pm  1.42$ \\
13	&	Starless Core	                          &  $1.32 \pm  0.10$ & -$0.91 \pm  0.13$ & $ 50.34 \pm 12.78$ \\
14	&	Free-Free Emission	                 &  $1.80 \pm  0.10$ &  $1.10 \pm  0.13$ & $ 17.02 \pm  1.10$ \\
15	&	Protostellar Core, Free-Free Emission     &  $1.35 \pm  0.10$ &  $1.02 \pm  0.10$ & $ 21.31 \pm  1.79$ \\
16	&	Starless Core	                          & $0.74 \pm  0.11$ & $ -0.30 \pm  0.2 $ & $-$ \\
17	&	Filament	                          &  $1.82 \pm  0.10$ &  $0.62 \pm  0.14$ & $ 19.76 \pm  1.53$ \\
18	&	Free-Free Emission	                  &  $1.47 \pm  0.10$ &  $0.89 \pm  0.10$ & $ 34.36 \pm  5.31$ \\
19	&	Protostellar Core, Free-Free Emission     &  $1.64 \pm  0.10$ &  $1.02 \pm  0.10$ & $ 24.78 \pm  2.50$ \\
20	&	Filament	                          &  $1.75 \pm  0.10$ &  $0.78 \pm  0.18$ & $ 21.59 \pm  1.87$ \\
21	&	Starless Core	                          &  $1.95 \pm  0.10$ &  $0.17 \pm  0.12$ & $ 19.08 \pm  1.43$ \\
22	&	Free-Free Emission	                  &  $1.65 \pm  0.10$ &  $0.35 \pm  0.10$ & $ 21.76 \pm  1.88$ \\
23	&	Protostellar Core	                  &  $1.65 \pm  0.10$ &  $1.27 \pm  0.12$ & $ 19.34 \pm  1.44$ \\
24	&	Filament	                          &  $0.97 \pm  0.10$ &  $0.61 \pm  0.12$ & $ 39.05 \pm  7.18$ \\ \hline
\end{tabular}
\caption{Fitted parameter values. The dust temperature $T_d$ in region 16 was not usefully constrained. }
\label{tbl:fitparams}
\end{center}
\label{tab:fitparams}
\end{table*}

\begin{table}
\begin{center}
\begin{tabular}{ |c|c|c|c| }
\hline
Marker Type & $\beta$ & $\beta'$ & $T$ [Kelvin] \\
\hline
Filament (N=5) & $1.77 \pm 0.04$ & $0.69 \pm  0.06$ & $15.2 \pm 0.5$  \\ 
Starless Core (N=5) & $1.55 \pm 0.04$ & $0.21 \pm 0.06$ & $15.8 \pm 0.6$ \\
Protostellar Core (N=10) & $1.62 \pm 0.03$ & $0.86 \pm 0.04$ & $17.4 \pm 0.4$ \\
Free-Free (N=8) & $1.48 \pm 0.04$ & $0.93 \pm 0.04$ & $20.4 \pm 0.6$ \\ \hline
All Regions (N=24) & $1.65 \pm 0.02$ & $0.74 \pm 0.02$ & $16.5 \pm 0.3$ \\
\hline
\end{tabular}
\caption{ Weighted average parameter values by physical characteristics of target region;  Slice 16 was excluded as a poorly-constrained outlier.  Error bars represent the error in the mean. }
\label{tbl:avgparams}
\end{center}
\label{tab:avebeta}
\end{table}

\subsubsection{Effect of Color Corrections}
\label{sec:colorcorr}

The models described in the previous sections are evaluated at the  response-weighted center frequency of each instrument's bandpass, as described in \S~\ref{subsec:bandpasses}. 
This is equivalent to approximating the spectrum as a mean plus a slope across the bandpass. To test the validity of this approximation 
we computed multiplicative corrections to the  model spectrum values:
\[
C_{i,j} = \frac{ <M_i(\nu)>_j}{M_i(\nu_{0,j})}
\]
where $M_i$ is the MBBall\_2beta best-fit model for region $i$,
$\nu_{0,j}$ are the  response-weighted frequencies for the band in question, and the average are over the $j$
bandpasses.   For the spectral average the applicable instrument
  bandpass is used. In effect these corrections convert the
  (approximate) monochromatic model predictions into bandpass-averaged
  model predictions, a conversion which is exact if the spectrum
  assumed to calculate $C_{i,j}$ is correct.  The largest corrections
  are for MAMBO (mean and RMS over regions: $1.08 \pm 0.02$),
  HERSCHEL-PACS $160 \, {\rm \mu m}$ ($0.97 \pm 0.02$), and GBT 1cm
  ($1.03 \pm 0.02$).  These three instruments have the largest
  fractional bandwidths in our analysis, with $\Delta \nu_{eff}/\nu_0
  \sim 40\% - 50\%$. The MBBall\_2beta fit was then repeated with the
  color-correction factors applied to the model predictions. The
  typical (median over regions) impact of the color correction on
  $\beta$ is $0.3 \sigma$ while for $\beta'$ it is $0.1 \sigma$, with
  the largest differences being $1.0 \sigma$ and $0.7\sigma$
  respectively. There is a modest shift in the median $T_d$ of $ - 0.3
  \sigma$, with the largest shift for an individual region being $-0.8 \sigma$. 

% note: the above works because what you really want to minimize is
%   (d - <m(nu)>)^2
% that is equivalent to minimizing
%  (<m>/m(nu))^2 * ( d * m(nu)/<m> - m(nu))^2
% for fixed <m>/m = 1/c

\vspace{0.3in}

\section{Interpretation}
\label{sec:interpretation}

\begin{figure*}
\centering
\includegraphics[width=\textwidth]{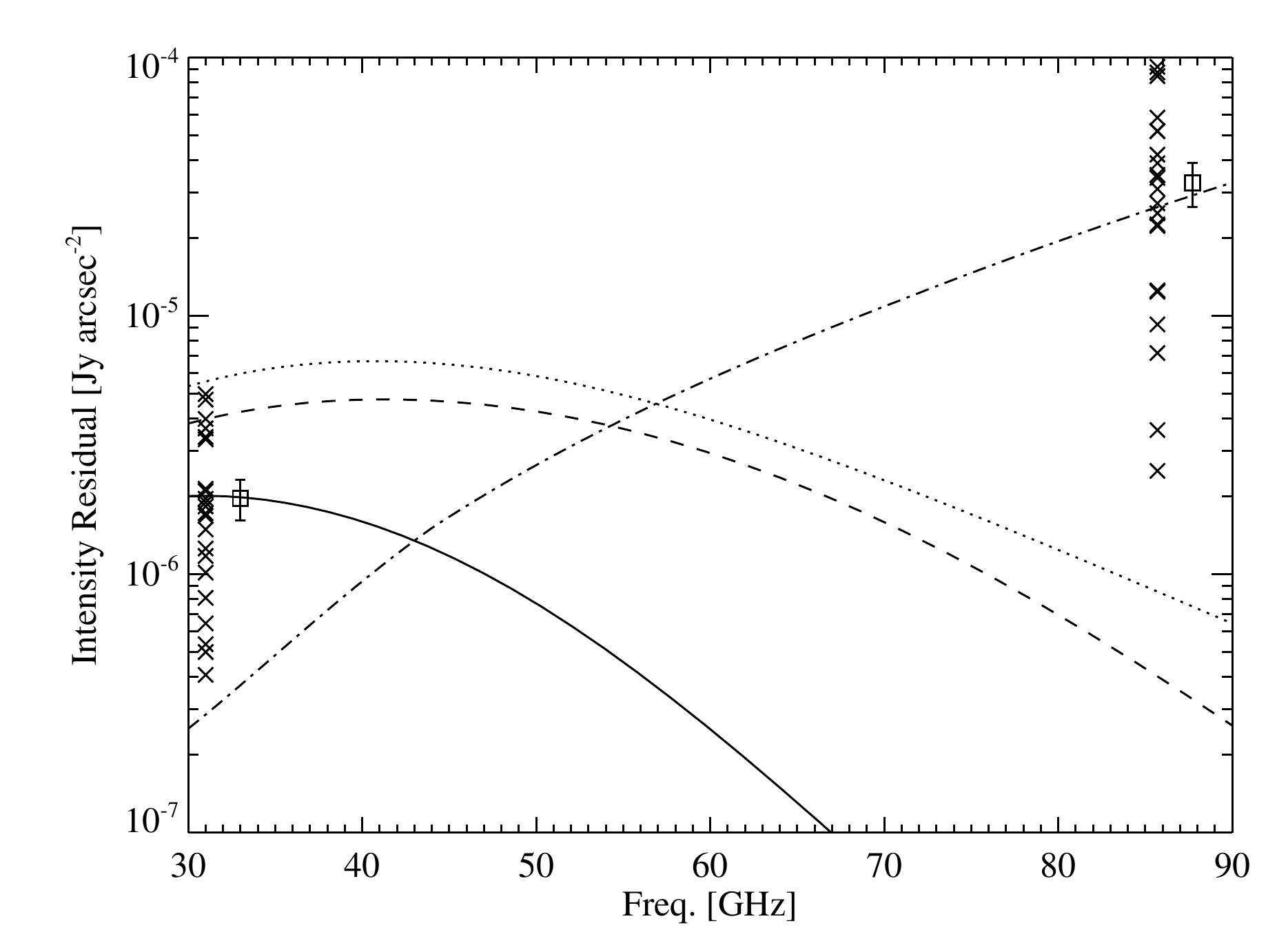}
\caption{Long wavelength residuals to the short wavelength fit (MBB2mm). Square points show the average and  error in the mean, with an overall $10\%$ calibration uncertainty added in quadrature. x's slightly offset in frequency are individual slice residuals. Three representative spinning dust spectra are shown for comparison. The solid line is the ``Dark Cloud'' model; the dashed line is the ``Molecular Cloud'' model; and the dotted line is the Dark Cloud model with density and temperature values appropriate for OMC 2/3 specifically.  The dash-dot line shows the spectrum of the $T_d = 17 \, {\rm K}$   model of \citet{paradis11} for dust grain emission including two-level system noise. }
\label{fig:residspec}
\end{figure*}

Our results indicate that a single modified black body SED provides a
considerably better description of the 2mm to 350~${\rm \mu m}$ SED
than it does when the GBT $3.3\, {\rm mm}$ and $1 \, {\rm cm}$ data
are included  ($\chi_n^2 = 1.65$ for $\lambda \le 2 \, {\rm mm}$ {\it
  vs.} $\chi_n^2 = 5.23$  when long wavelength data are
included). Similar results were obtained by S16.  Structures at 3~mm
and 1~cm are seen to be significantly brighter than what would be
expected by extrapolating the short wavelength MBB SED; and the
overall SEDs are better described by a ``broken'' blackbody with a
different dust opacity index $\beta'$ at $\nu < 120 \, {\rm GHz}$,
than they are a standard single-$\beta$ MBB (typical $\chi_n^2 = 2.98$
vs $5.23$).  The long-wavelength residuals to a standard MBB fit to
the $\lambda < 3 \, {\rm mm}$ data are shown in
Figure~\ref{fig:residspec}. As noted in S14, the morphological
similarity of the MUSTANG-2 maps with the images at shorter
wavelengths is a strong indication the 3~mm emission we see is
dominated by dust, or by something closely associated with dust in the
ISM. In the next three subsections we consider several possible
explanations for the observed spectrum of dust-associated emission.

\subsection{Spectral Line Contamination}
\label{sec:linecontamination}

Molecular line emission could in principle contaminate the broad-band continuum SEDs we have measured. Transitions of the brightest molecular species in the millimeter regime, CO, lie  outside the MUSTANG-2 and GISMO bandpasses. Although CO(2-1) is within the MAMBO bandpass, S14 argue that it is unlikely to significantly contaminate the 1mm continuum measurements. To the extent it does it would tend to steepen 1mm-3mm spectra rather than flatten them, and would not contribute to the 1cm and 3mm flattening we report.

% SPECULATIOn REMOVED -- 
%We note that this could be a factor contributing to the outliers detected at 1mm in slices 11 and 13 (\S~\ref{subsec:slices}) %if there are particularly intense outflows with broad velocity profiles. Several other 1mm points show indications of a lower %level, positive bias.

It is possible that weaker lines could contribute to the measured 3mm continuum level.  Because MUSTANG and MUSTANG-2 have different bandpasses ($80$-$99 \, {\rm GHz}$ {\it vs.} $75$-$105 \, {\rm GHz}$), the excellent agreement between the original MUSTANG maps of S14 and those presented here suggests that that is not the case.  To further check this possibility of line contamination we used high spectral-resolution ALMA Compact Array (ACA) data from project {\tt 2013.1.00662.S} (\S~\ref{subsec:otherdata}) to search for line contamination. These data covered OMC-2 and OMC-3 over a narrow ($91.2$ - $91.7$ GHz) frequency range at a spectral resolution of $\sim 35 \, {\rm kHz}$.  While several spectral lines are evident (e.g. ${\rm CH_3 CN}$), their excision changes the integrated continuum fluxes by $<2\%$. Furthermore the ACA-observed source fluxes agree with the MUSTANG-2 source fluxes within a few percent.  We conclude that contamination by spectral lines is unlikely to account for any significant portion of the elevated 3mm emission we report.

%We also compared the MUSTANG-2 image to the publicly available ALMA 7m+12m array observations of OMC 2/3 \citep[][see also \S~\ref{subsec:otherdata}]{Kainulainen2017} and found that for compact sources the emission agrees within a few percent.  These data also sample a different frequency range (ranging from 98 to 108 GHz in four individual 2~GHz wide spectral windows).  We also re-imaged the ALMA 7-m array data excluding the continuum channels most likely to have bright line emission and found that the 3mm intensities changed by $< 2\%$.  We conclude that contamination by spectral lines is unlikely to account for any significant portion of the elevated 3mm emission we report.

\subsection{Free-free \& Synchrotron Emission}

Thermal Bremsstrahlung (``free-free'') emission is typically the dominant emission mechanism in galaxies in the ``trough'' between synchrotron and
thermal dust emission, 10 GHz to 100 GHz or so. In regions of high-mass star formation, such as OMC 1,
thermal free-free is even more prevalent.  OMC 2/3 itself does not harbor massive stars or star clusters as does OMC 1, although
as previously noted there is an HII region (M43) created by an O/B star immediately to the South-East.  Young protostars can 
give rise to free-free emission, and in fact the $8.3$~GHz VLA maps of R99 reveal 14 sources in OMC 2/3 with
flux densities between $0.15 \, {\rm mJy}$ and $2.84 \, {\rm mJy}$.  The $8''$ beam of these observations is well matched to
the scales being studied here, so these data provide a good handle on the contribution of free-free emission to
the SEDs of the regions under study.   As discussed in \S~\ref{sec:1cm}, the 31 GHz (1~cm) data reported here have been
corrected for free-free using the R99 measurements under the assumption that each source is optically thin at $8.3$~GHz.
The contribution of optically thin free-free to the $3.3$~mm (90 GHz) points is negligible in all regions except one, Slice 14, where it explains $\sim 25\% $ of the peak  $3.3$~mm surface
brightness at the location in question.

%
% CORRECT SLICE 14 @ 3MM  ?? (the only free-free that matters)
%

While optically thin free-free sources cannot explain any significant portion of the long wavelength spectral residuals it is possible that 
free-free which is optically thick over at least part of the frequency range measured could contribute to the observed
spectral signal.  Such a signal would be spatially unresolved at the resolutions achieved in our data, and have a flux density increasing
as $\nu^2$ up to a turn-over frequency somewhere between $8.3 \, {\rm GHz}$ (the frequency of the VLA maps in R99) and $100 \, {\rm GHz}$ or so.
In order to account for any significant fraction of the 90 GHz spectral excess the turn-over frequency would have to be above 31 GHz in most cases.
To test the potential relevance of this mechanism we examined the flux density ratios between $8.3$, $31$ and $90 \, {\rm GHz}$ for the 6
regions which have $8.3 \, {\rm GHz}$ free-free sources closely associated with 90 GHz emission.  Optically thick free-free
with a turnover well above 31~Ghz would give $S_{31}/S_{8.3}= (31/8.3)^2 = 13.8$.  Five of the six sources with $8.3$~GHz detections have $S_{31} / S_{8.3}$ 
lower than this by a factor of 2 to 7.  The remaining region, slice 15, has $S_{31}/S_{8.3} = 11.25 \pm 0.19$, and thus, the 90 GHz and 31 GHz measurements could be explained by free-free emission with a turn-over between $8.3 \, {\rm GHz}$ and $31 \, {\rm GHz}$. Were this to be the case, the optically thick free-free still does not contribute significantly to the observed 3mm spectral residual since the turn-over frequency is too low for it to do so.

Synchrotron radiation does not typically track dust spatially. Dust is
prevalent in regions where the ISM is cold, dense, and heavily
shielded from ionizing radiation. Synchrotron radiation requires a
source of relativistic electrons such as an active galactic nucleus or
supernova.  If the 31~GHz and 90~GHz signals reported here were
generated by diffuse synchrotron or free-free radiation, these
structures would be clearly visible in the $8.3$~GHz maps of
\citet{Reipurth1999} and even more so in the NVSS.  This not being the
case, synchrotron is clearly not a significant contributor to the SED
measurements we report.  We can place an upper limit on the surface
brightness of synchrotron or free-free radiation using the NVSS, which
has a $45''$ (FWHM) beam and a typical sensitivity of $0.45 \, {\rm
  mJy/bm}$, or a surface brightness sensitivity of $2.0 \times 10^{-7}
\, \rm Jy \, arcsec^{-2}$. For an optically thin free-free spectrum,
this would give to a $95\%$ upper limit of $2.8 \times 10^{-7} \, \rm
Jy \, arcsec^{-2}$ at 31 GHz and $2.6 \times 10^{-7} \, \rm Jy \,
arcsec^{-2}$ at 90 GHz. The corresponding limits on synchrotron are
$4.4 \times 10^{-8} \, \rm Jy \, arcsec^{-2}$ (31 GHz) and $2.2 \times
10^{-8} \, \rm Jy \, arcsec^{-2}$ (90 GHz).  The limits from
\citet{Reipurth1999} are weaker by a factor of $\sim 3$ (free-free) to
$8$ (synchrotron) due to its lower surface brightness sensitivity and
closer proximity in frequency.

\subsection{Anomalous Microwave Emission}

High surface brightness observations in the microwave regime over the past two decades have 
revealed the widespread presence of dust-associated emission which is not well described by thermal dust emission, thermal bremstrahlung (free-free), or synchrotron radiation \citep{kogut1996,leitch1997,deoliveiracosta1997}. This so-called ``Anomalous Microwave Emission'' (AME) is best explained as electric dipole emission from small, charged spinning dust grains  \citep{draine1998}, although there are other plausible models such as magnetic dipole emissions from large, ferrous dust grains \citep{draine1999} or spinning nano-silicates \citep{hoang2016}, and recent observational evidence suggests that nano-diamonds could be relevant in some environments \citep{greaves}. 
Although most studies of AME target our own Galaxy, it has also been detected in other, nearby galaxies \citep{murphy2010,murphy2018}. For a recent review of the state of the art in AME observations and theory see \citet{dickinson}.

AME spectra are generally seen to peak in the 20 - 60 GHz range, with the peak frequency and detailed spectral shape depending on the local physical conditions including the grain size distribution, the intensity of the radiation field, the dipole moment of the emitting grains, the number density of grains, and the ionization fractions of hydrogen and carbon \citep{finkbeiner2004,planckXX2011,planckDust2015}.  We  used the SPDUST2 code \citep{alihamoud2009,silsbee2011} to compute
representative AME spectra. Figure~\ref{fig:residspec} shows spectra for diffuse molecular cloud (MC) and dark cloud (DC) environments, with parameters as in \citet{alihamoud2009}, as well as the spectrum for a DC environment with density and temperature set to values closer to those observed in the OMC 2/3 filament ($n_H = 10^5 \, {\rm cm^{-3}}$ and $T_d = 16 \, {\rm K}$). For comparison the residuals to the short wavelength MBB fit (MBB2mm) of the SEDs the 24 individual regions are also shown. The AME models do not match the data well.

It is interesting to note that, as illustrated by Fig.~5 of \citet{ysard2011}, standard AME spectra can peak at frequencies as high as 100~GHz or even higher when the interstellar radiation field (ISRF) is strong, $G_0 \gtrsim 5$.  While there are several significant sources of ionizing photons nearby--- particularly Nu Ori, as well as the Trapezium cluster slightly farther afield--- the abundant dust in this very dense MC environment will attenuate these photons over a relatively short distance \citep[although see][]{jorgensen2006}.  When \citet{ysard2011} compute line-of-sight integrated SEDs for dense molecular cloud structures using radiative transfer with physically plausible density and temperature distributions for the MC material,  AME spectra with peaks at $\nu < 50 \, {\rm GHz}$ are obtained.

%in the form of small spinning dust grains has been shown to be an important component of galactic %dust emission,  \citep{dickinson}. \citet{planckAME} identify 27 AME regions with spectra consistent %with model spectra of spinning dust. The canonical model of a ``Cold Neutral Medium'' can produce %emission from spinning dust at 3.3~mm at a level of $1.05 \times 10^{-18} Jy/Sr/(H/cm^2)$, which %%corresponds to a level of 23 $\mu$Jy/beam, however this assumes densities of 30 hydrogen %atoms/$cm^3$, this is much lower than the densities $~10^5 cm^{-3}$ observed in OMC2/3 (see S14 and %S16). Using this value for the density pushes the peak emission to higher frequencies (~275 GHz) and %would overestimate the observed emission.

% density ~1e5 /cm3 (Johnstone & Bally 1999) - per s14
% also cite draine & weingartner 01, tibbs+15, & Ali-Haimoud+ 09

\subsection{Dust Emission}
\label{subsec:dust}

An alternate explanation for the higher emissivity we see at 3~mm and 1~cm is that the
dust itself has intrinsically higher emissivity at long wavelength compared to what
would be expected based on the MBB extrapolation from shorter wavelength data. An attractive
feature of this model is that it naturally explains the excellent correspondence seen
between the morphology of the 3~mm map and the shorter wavelength maps. It also explains 
the presence of enhanced emissivity in the full range of environments studied here--- optically thick 
free-free for instance cannot explain the long wavelength excess that is observed in the filament
itself.  

Intriguingly, the {\it Planck} data shown in Figure~3 of S14 hints at a flattening spectrum  at $\lambda > 4 \, {\rm mm}$ similar to that seen here. The evidence
is inconclusive owing to {\it Planck}'s very different spatial frequency
filtering and low angular resolution
compared to the scale of the structures being studied: $10' - 33'$, for {\it Planck} measurements between 100 GHz to 31 GHz respectively, as opposed to the $25''$-$3'$ scales studied here. 
The effects of beam dilution on the SED derived from {\it Planck} data
will also vary considerably with frequency over the range of interest. The {\it Planck} collaboration \citep{planck13XI} do find that  a single $\beta$ modified blackbody fit from 353 to 3000~GHz, extrapolated down to 100~GHz, also under predicts the observed {\it Planck} dust emission. \citet{MeisnerFinkbeiner} fit this data to a two-component grey-body model and are able to accurately predict the 100GHz emission.  The measurements reported here suggest that this flattened spectrum continues to a wavelength of 1~cm.

Detailed models of emission from amorphous dust grains based on laboratory measurements
predict a flattened spectrum (higher emissivity) at long wavelength \citep{meny07,coupeaud11,paradis11}, mainly due to two-level system fluctuations in the dust grains. Figure~\ref{fig:residspec} shows that while this model could help explain the 3mm data points, it under-predicts the 1cm emissivity.   Thus, multiple factors could be relevant to the data we present here.

%{\bf note: the beta value directly between the spectral excess data points is about 0.4. vs 0.7, the imperfect fit given a 120 GHz kink.}

\section{Conclusions}
\label{sec:conclusions}

We have presented sensitive, new 3~mm and 1~cm continuum measurements
of the ``Integral-shaped Filament'', OMC 2/3.  These measurements
confirm the enhanced 3~mm brightness of OMC 2/3 first reported in
\citet{schnee2014}, and support similar findings at lower resolution
throughout the Galaxy as a whole
\citep[e.g.][]{planck13XI,MeisnerFinkbeiner}.  Our data indicate that
this finding extends to very dense MC environments and potentially to
frequencies as low as 30 GHz.  Consistent with \citet{sadavoy2016}, we
find that the data from 2~mm to $160 \, {\rm \mu m}$ are in contrast
well described by a simple modified black body with $\beta \sim
1.6$. This supports the suggestion advanced by S16 that there may be a
deviation from a simple modified blackbody SED at long wavelength. If
the long-wavelength data are empirically modeled by allowing a break
in the spectrum at $120 \, {\rm GHz}$, the 1~cm-3~mm spectrum
corresponds to $\beta' \sim 0.7$.  The long-wavelength emissivity
index $\beta'$ is generally consistent in most of the regions studied
(filament, protostellar core, and regions with a known free-free
source), but significantly flatter in the five starless cores studied
here, with $\beta' = 0.21 \pm 0.06$.

Some models of radiation from amorphous dust grains do predict
enhanced long wavelength brightness \citep{meny07, coupeaud11,
  paradis11}. This explanation would naturally explains the excellent
correspondence of the 3~mm continuum maps with the morphology seen at
shorter wavelengths.  While this model may explain much of the
observed 3~mm brightness it appears to under-predict the 1cm signal.
Canonical AME models do not provide a good fit to the observed long
wavelength spectrum, but could account for the 1cm signal.  In some of
our target regions, particularly those with known protostars, we
cannot rule out a contribution from thermal Bremsstrahlung which is
optically thick up to a turnover frequency at $\nu > 31 \, {\rm
  GHz}$. The spectra of {\it known} free-free regions in OMC 2/3,
however, provides no evidence for such a scenario.  Free-free cannot
account for the observed long-wavelength SED in the filament itself,
and is unlikely to do so in starless cores.

In spite of the relative faintness of dust emission at 3~mm, interferometric surveys of known star forming regions are common at this wavelength \citep[e.g.][]{carpenter2002,eisner2006,dunham2016,kirk2017,Kainulainen2017}. In part this is due to the relatively large fields of view and higher instrumental sensitivity that interferometers provide at these lower frequencies.  The data from such surveys are also less affected by optical depth considerations than shorter wavelength data.   Our findings could have substantial implications on the interpretation of data from 3~mm core surveys, suggesting that masses and densities of cores determined from them could be overestimated by a factor of 2--3. Such an overestimate would correspondingly impact dynamical stability estimates of the cores made from 3~mm measurements.  In order to interpret these surveys reliably, and  to conclusively understand the physical conditions in this region and regions like it, high quality observations at intermediate and lower frequencies will be needed as well as more detailed modeling of the dust and ISM conditions.

\acknowledgments {\bf Acknowledgments:}  Sara Stanchfield was supported by NASA NSTRF NNX14AN63H. Other investigators at the University of Pennsylvania
were supported by NSF grant 1615604, and by the Mt. Cuba Astronomical Foundation. 
We thank Diego Mardones, Di Li, and Jeremy Ren for sharing their ALMA data;  Scott Schnee, Dana Balser, Loren Anderson, and Remy Indebetouw for insightful comments about the scientific interpretation of the data we present;  Luca Di Mascolo for expert advice  on AplPy;  Johannes Staguhn for providing the GISMO bandpass; and Bryan Butler for providing the VLA data on 3C138 which facilitated calibrating our Ka-band GBT data. The National Radio Astronomy Observatory and the Green Bank Observatory are facilities of the National Science Foundation operated under cooperative agreement by Associated Universities, Inc.  This paper makes use of the following ALMA data: ADS/JAO.ALMA\#2013.0.00662.S. ALMA is a partnership of ESO (representing its member states), NSF (USA) and NINS (Japan), together with NRC (Canada), MOST and ASIAA (Taiwan), and KASI (Republic of Korea), in cooperation with the Republic of Chile. The Joint ALMA Observatory is operated by ESO, AUI/NRAO and NAOJ.

\facilities{GBT, VLA}

%\facility{facility ID}
%\facilities{facility ID, facility ID, facility ID} 

\software{AplPy \citep{aplpy},
Astropy \citep{astropy},
CASA \citep{casa},
ds9 \citep{ds9}}

\bibliographystyle{yahapj}
\bibliography{accepted_ms}

\end{document}